\documentstyle[amssymb,prl,aps,epsf,graphicx,floats]{revtex}

\begin{document}
\draft

\twocolumn[\hsize\textwidth\columnwidth\hsize\csname@twocolumnfalse\endcsname
\title{Trajectory structures and transport }
\author{Madalina Vlad, Florin\ Spineanu}
\address{National Institute of Fusion Science, Toki 509-5292, Japan\\
and\\
National Institute for Laser, Plasma and Radiation Physics, \\
Association Euratom-MEC Romania, P.O.Box MG-36, Magurele, Bucharest, Romania\\
}
\maketitle

\begin{abstract}
The special problem of transport in 2-dimensional divergence-free stochastic velocity fields is studied by developing a statistical approach, the nested subensemble method. The nonlinear process of trapping determined by such fields generates trajectory structures whose statistical characteristics are determined. These structures strongly influence the transport. 
\end{abstract}


\twocolumn]\narrowtext

\section{Introduction}

Test particle motion in stochastic velocity fields is a generic problem in
various topics of fluid and plasma turbulence or solid state physics \cite
{mccomb}-\cite{K02}. This problem was very much studied and several
analytical approaches were developed. Most of them are based on Corrsin \cite
{Corrsin}, \cite{mccomb} and direct-interaction \cite{Roberts}, \cite{BPP}
approximations.

In this context, particle motion in 2-dimensional divergence-free velocity
fields represents a special case. Kraichnan has shown for the first time, in
a study based on numerical simulations \cite{kraichnan}, that the existing
analytical methods are not adequate for this type of problems. The cause of
this anomaly is the trapping of the particles, which appears in such
velocity fields when they have slow time variation. The trapping consists in
trajectory winding on almost closed paths. A typical trajectory has a
complicated shape with such localized trapping events separated by long
jumps. Consequently, the probability distribution function is non-Gaussian.
Thus Corrsin and direct-interaction approximation which are based on this
hypothesis are not adequate for this specific case. A more recent analysis
of the effects of trapping is presented in Ref.\cite{Physicalia} where a
non-Gaussian peaked distribution of the displacements and a long negative
tail in the Lagrangian velocity correlations are evidenced for numerically
calculated trajectories.\ 

This special problem of diffusion in 2-dimensional divergence-free velocity
fields describes for instance the transport in turbulent magnetized plasmas
or in incompressible fluids. It was studied especially by means of direct
numerical simulations (\cite{jacques} and the reference there in) or on the
basis of simplified models \cite{majda}, \cite{Pecs}. There is also a
qualitative theoretical estimation of the scale law for the asymptotic
diffusion coefficient \cite{isichenko} based on an analogy with the
percolation process in stochastic landscapes. The case of collisional
particle motion in such static velocity fields was analyzed by means of the
renormalization group techniques (\cite{BG} and the references there in) and
the asymptotic time behavior of the mean square displacement was determined.
The evolution of the diffusion process is determined only in \cite{VSMB1}
where a statistical approach, the decorrelation trajectory method, is
developed. It yields analytical expressions for the time dependent diffusion
coefficient $D(t)$ and for the correlation of the Lagrangian velocity $L(t),$
that are qualitatively valid for the whole range of the Kubo number (see
next Section for the definitions). The basic idea consists of determining
the Lagrangian velocity correlation by means of a set of average Lagrangian
velocities estimated in subensembles of realizations of the stochastic
field. This method could be extended to more complicated physical systems
which contain particle collisions \cite{VSMB2}, average velocities \cite
{VSMB3} or a supplementary component of the motion perpendicular to the
2-dimensional plane \cite{NF}, \cite{PRE03}. It was shown that the presence
of trapping strongly influences the diffusion coefficients and their scaling
laws determining a rich class of anomalous diffusion regimes. These studies
have shown that the decorrelation trajectory method provides a qualitatively
good description of the trapping process. However, due to the rather strong
approximation introduced in this method (see Section III), there are several
qualitative aspects that are not well described \cite{Com}, \cite{Repl}.
They are related to trajectory fluctuations and their correlation with the
stochastic velocity.

The above results concern the effect of trapping on the individual
trajectories. The trapping has also collective effects. It determines
coherence in the stochastic motion in the sense that bundles of neighboring
trajectories form localized structures similar with fluid vortices. The
formation of these structures strongly influences the transport. The aim of
this paper is to study the statistical characteristics of these trajectory
structures determined by the intrinsic trapping appearing in 2-dimensional
divergence-free stochastic velocity fields. The average, the dispersion and
the probability distribution function for the trajectories in such
structures are determined. The evolution of these statistical quantities
saturates showing that they do not contribute to the asymptotic transport.
The statistical evolution of the distance between two neighboring
trajectories is also studied and its mean, dispersion and probability
distribution are determined. We show that for the trapped trajectories in a
static velocity field, the average decays to zero and the dispersion remains
very small for long time and eventually saturates. A very strong clump
effect is found for the trapped trajectories. This demonstrates the
existence of trajectory structures and their effect of strongly reducing the
relative transport. The later is produced only by a small part of the
stochastic trajectories which are not contained in these vortical trajectory
structures.

The method developed for this study is a semi-analytical statistical
approach which extends and improves the decorrelation trajectory method \cite
{VSMB1} by introducing the fluctuations of the trajectories in the
subensembles.

\section{The problem}

Particle motion in a 2-dimensional stochastic velocity field is described by
the nonlinear Langevin equation: 
\begin{equation}
\frac{d{\bf x}(t)}{dt}={\bf v}\left[ {\bf x}(t),t\right] ,\qquad {\bf x}(0)=%
{\bf 0}  \label{1}
\end{equation}
where ${\bf x}(t)$ represents the trajectory in Cartesian coordinates ${\bf x%
}\equiv (x_{1},x_{2}).$ The stochastic velocity field ${\bf v}({\bf x},t)$
is \ divergence-free: ${\bf \nabla \cdot v}({\bf x},t)=0$ and thus its two
components $v_{1}$ an $v_{2}$ can be determined from a stochastic scalar
field $\phi ({\bf x},t)$, as: 
\begin{equation}
v_{i}({\bf x},t)=\varepsilon _{ij}\frac{\partial \phi ({\bf x},t)}{\partial
x_{j}}  \label{vd}
\end{equation}
where $\varepsilon _{ij}$ is the antisymmetric tensor ($\varepsilon _{12}=1,$
$\varepsilon _{21}=-1,$ $\varepsilon _{11}=\varepsilon _{22}=0$). In the
studies of turbulence of magnetized plasmas, $\phi ({\bf x},t)$ is
essentially the potential ($\phi =-\phi ^{e}/B$ where $\phi ^{e}({\bf x},t)$
is the electrostatic potential and $B$ is the magnetic field strength) and
in fluid turbulence $\phi ({\bf x},t){\bf e}_{3}$\ is the stream function ($%
{\bf e}_{3}$ is the unitary vector along the axis perpendicular on the plane 
$(x_{1},x_{2})).$

The potential $\phi ({\bf x},t)$ is considered to be a stationary and
homogeneous Gaussian stochastic field, with zero average and given two-point
Eulerian correlation function (EC) 
\begin{equation}
E({\bf x},t)\equiv \left\langle \phi ({\bf x}_{1},t_{1})\,\phi ({\bf x}_{1}+%
{\bf x},t_{1}+t)\right\rangle  \label{pec}
\end{equation}
where $\left\langle ...\right\rangle $ denotes the statistical average over
the realizations of $\phi ({\bf x},t).$ The statistical properties of the
space derivatives of the potential are completely determined by those of the
potential. They are stationary and homogeneous Gaussian stochastic fields
like $\phi ({\bf x},t)$. The two-point Eulerian correlations of the
derivatives of $\phi ({\bf x},t)$ are obtained as derivatives of the
potential EC, $E({\bf x},t).$ We introduce the notation $E_{i..;k..}({\bf x}%
,t)\equiv \left\langle \phi _{i..}({\bf x}_{1},t_{1})\,\,\phi _{k..}({\bf x}%
_{1}+{\bf x},t_{1}+t)\right\rangle $ where $\phi _{i}({\bf x},t)\,\equiv 
\frac{\partial }{\partial x_{i}}\phi ({\bf x},t)$ and the subscript of $E$\
contains the indices of the derivatives of the potential in ${\bf x}_{1},$ $%
t_{1}$ (left factor) separated by semicolon \ from the indices of the
derivatives of the potential in ${\bf x}_{1}+{\bf x},$ $t_{1}+t$ (right
factor). The absence of indices correspond to a factor $\phi $\ inside the
average (see Eqs. (\ref{cev}) for examples). One obtains 
\begin{equation}
E_{i..;k..}({\bf x},t)=(-1)^{n}\frac{\partial }{\partial x_{i}}...\frac{%
\partial }{\partial x_{k}}...E({\bf x},t)  \label{ecder}
\end{equation}
with $n$ equal to the number of derivatives of the first factor $\phi _{i..}(%
{\bf x}_{1},t_{1})$ inside the above average. In particular, the velocity $%
{\bf v}({\bf x},t)$ is such a stationary and homogeneous Gaussian stochastic
field. The correlation of the velocity components and \ the
potential-velocity correlations are obtained using the definition (\ref{vd})
of the velocity and Eq. (\ref{ecder}) as 
\begin{eqnarray}
\left\langle v_{1}({\bf 0},0)\,v_{1}({\bf x},t)\right\rangle &=&E_{2;2}({\bf %
x},t)=-\frac{\partial ^{2}}{\partial x_{2}^{2}}E({\bf x},t),  \label{cev} \\
\left\langle v_{2}({\bf 0},0)\,v_{2}({\bf x},t)\right\rangle &=&E_{1;1}({\bf %
x},t)=-\frac{\partial ^{2}}{\partial x_{1}^{2}}E({\bf x},t),  \nonumber \\
\left\langle v_{1}({\bf 0},0)\,v_{2}({\bf x},t)\right\rangle &=&-E_{2;1}(%
{\bf x},t)=\frac{\partial ^{2}}{\partial x_{1}\partial x_{2}}E({\bf x},t), 
\nonumber \\
\left\langle v_{i}({\bf 0},0)\phi ({\bf x},t)\,\right\rangle &=&\varepsilon
_{ij}E_{j;}({\bf x},t)=-\varepsilon _{ij}\frac{\partial }{\partial x_{j}}E(%
{\bf x},t),  \nonumber \\
\left\langle \phi ({\bf 0},0)v_{i}({\bf x},t)\,\right\rangle &=&\varepsilon
_{ij}E_{;j}({\bf x},t)=\varepsilon _{ij}\frac{\partial }{\partial x_{j}}E(%
{\bf x},t).  \nonumber
\end{eqnarray}
These correlations (\ref{ecder})-(\ref{cev}) will be used in the following
calculations.

The potential is a continuous function of ${\bf x}$ and $t$ in each
realization and it determines an unique trajectory as the solution Eq. (\ref
{1}). Starting from the above statistical description of the stochastic
potential and from an explicit EC, $E({\bf x},t),$ one has to determine the
statistical properties of the trajectories. The later can be obtained from
the Lagrangian velocity correlation (LVC),\ defined by:

\begin{equation}
L_{ij}(t)\equiv \left\langle v_{i}\left[ {\bf x}(0),0\right] )v_{j}\left[ 
{\bf x}(t),t\right] \right\rangle  \label{CL}
\end{equation}
for a stationary process. The mean square displacement $\left\langle
x_{i}^{2}(t)\right\rangle $ and its derivative, the running diffusion
coefficient $D_{i}(t)$, are determined by this function \cite{Taylor} as: 
\begin{equation}
\left\langle x_{i}^{2}(t)\right\rangle =2\int_{0}^{t}d\tau \;L_{ii}(\tau
)\;(t-\tau ),  \label{MSD}
\end{equation}
\begin{equation}
D_{i}(t)=\int_{0}^{t}d\tau \;L_{ii}(\tau ).  \label{D}
\end{equation}
The probability distribution function of the trajectories can be obtained
from the equation for particle density, once the LVC is known \cite{B}.
Thus, the solution of the above Langevin problem consists essentially in
determining the LVC (\ref{CL}) corresponding to a given EC (\ref{pec}) of
the stochastic potential.

This kind of Langevin problem, sometimes named in the literature diffusion
by continuous movements, is nonlinear due to the space dependence of the
potential, which leads to ${\bf x}$-dependence of the EC (\ref{pec}). The
importance of the nonlinearity is characterized by the Kubo number defined by

\begin{equation}
K=\frac{V\tau _{c}}{\lambda _{c}}\,  \label{K}
\end{equation}
where $V$ is the amplitude of the stochastic velocity, $\tau _{c}$ is the
correlation time and $\lambda _{c}$ is the correlation length. These
parameters appear in the EC of the velocity as the maximum value in the
origin [$V^{2}=E_{ii}({\bf 0},0)]$ and the characteristic decay time and
length of this functions. The Kubo number is thus the ratio of $\tau _{c}$
to the average time of flight of the particles over the correlation length, $%
\tau _{fl}=\lambda _{c}/V.$ \ It measures the particle's capacity of
exploring the space structure of the stochastic velocity field before it
changes.

For small Kubo numbers the time variation of the velocity field is fast and
the particles cannot ''see'' the space structure of the velocity field. The
condition $K\ll 1$ $(\tau _{c}\ll \tau _{fl})$ defines the quasilinear
regime (or the weak turbulence case) for which the results are well
established: the diffusion coefficient is $D_{ql}=($ $\lambda _{c}^{2}/\tau
_{c})K^{2}$ and the trajectories have Gaussian distribution.

For $K>1$ ($\tau _{c}>\tau _{fl}$) the time variation of the stochastic
potential is slow and the trajectories approximately follow the contour
lines of $\phi ({\bf x},t).$ This produces a trapping effect : the
trajectories are confined for long periods in small regions.\ A typical
trajectory shows an alternation of large displacements and trapping events.
The latter appear when the particles are close to the maxima or minima of
the potential and consist of trajectory winding on almost closed small size
paths. The large displacements are produced when the trajectories are at
small absolute values of the potential. Thus there is a strong influence of
trapping on individual trajectories. Trajectory trapping appears for $K>1$
and becomes stronger as $K$ increases up to the limit of static field ($K,$ $%
\tau _{c}=\infty )$ where the trapping is permanent. It determines the
decrease of the diffusion coefficient and the \ change of its dependence on
the Kubo number from the Bohm scaling \cite{Dupree}, \cite{TMcN}, $D_{B}\sim
($ $\lambda _{c}^{2}/\tau _{c})K,$ to a trapping scaling, $\ D_{tr}\sim ($ $%
\lambda _{c}^{2}/\tau _{c})K^{\gamma }$ with $\gamma <1.$ This process was
statistically described in \cite{VSMB1}, \cite{VSMB2} and the exponent $%
\gamma $ was evaluated for given EC of the potential. In the limit of static
potential field ($K,$ $\tau _{c}=\infty )$ the transport is subdiffusive
with $D(t)\rightarrow 0$ and $\left\langle x^{2}(t)\right\rangle \sim
t^{\alpha }$ $\ $with $\alpha <1$. It can be shown (see Section V) that $%
\alpha =\gamma $ and thus the same power law describes the large time
behavior of the mean square displacement in a static potential and the
dependence on $K$ of the asymptotic diffusion coefficient in a time
dependent potential.

Besides this influence on individual trajectories, the trapping has
collective effects and it generates trajectory structures similar with fluid
vortices. The trapping appears coherently for bundles of neighboring
trajectories leading to eddying regions. We analyze here the statistical
characteristics of these trajectory structures. We show that the dispersion
of the trapped trajectories saturates and that the mean square of the
distance between trajectories evolve slowly and eventually saturates. The
motion of trapped particles is almost coherent and leads to structures which
do not contribute to the transport (trajectory dispersion) nor to the
relative transport.

\section{The nested subensemble method}

We start from the main idea of the decorrelation trajectory method \cite
{VSMB1}. It consists in studying the Langevin equation (\ref{1}) in
subensembles (S) of realizations of the stochastic field, which are
determined by given values of the potential and of the velocity in the
starting point of the trajectories: 
\begin{equation}
(S):\quad \phi ({\bf 0},0)=\phi ^{0},\quad {\bf v}({\bf 0},0)={\bf v}^{0}.
\label{2}
\end{equation}

We note that similar subensemble averages of the Eulerian stochastic
velocity field were studied in Ref.\cite{Adrian} with the aim of showing
that eddies and structures exist even in isotropic turbulence. Subensemble
Lagrangian averages are estimated in \cite{Philip} on the basis of a rather
strong assumption and in \cite{Pecs} for a model of rotating fluid layers.
Our approach is different.\qquad

The statistical properties of the stochastic potential and velocity, reduced
in the subensemble (S) defined by condition (\ref{2}) are derived in \cite
{VSMB1}. They are Gaussian fields but non-stationary and non-homogeneous,
with space and time dependent averages and dispersions. The averages depend
on the parameters of the subensemble and are defined by 
\begin{equation}
\Phi ^{E}({\bf x,}t;S)\equiv \left\langle \phi ({\bf x},t)\right\rangle
_{S},\quad {\bf V}^{E}({\bf x,}t;S)\equiv \left\langle {\bf v}\left[ {\bf x}%
,t\right] \right\rangle _{S}  \label{vfims}
\end{equation}
where $\left\langle ...\right\rangle _{S}$ denotes the average taken on the
realizations in (S) and the superscript $E$ is used to underline the
Eulerian nature of these quantities. They are equal to the corresponding
imposed condition (\ref{2}) in ${\bf x=0}$ and $t=0$ and decay to zero at
large distance and/or time. The mean squares of the potential and velocity
fluctuations are zero in ${\bf x=0}$, $t=0$ and increase up to the values
corresponding to the whole set of realizations at large distance and/or
time. The existence of an average Eulerian velocity in the subensemble
determines an average motion (decorrelation trajectory). This decorrelation
trajectory is estimated in each subensemble and an approximation for the LVC
is derived in terms of the set of these trajectories.

More precisely, the LVC (\ref{CL}) for the whole set of realizations is
obtained by summing up the contributions of each subensemble. The latter can
be written as $\left\langle v_{i}\left[ {\bf x}(0),0\right] \,v_{j}\left[ 
{\bf x}(t),t\right] \right\rangle _{S}=v_{i}^{0}\,\left\langle v_{j}\left[ 
{\bf x}(t),t\right] \right\rangle _{S}$. Thus, the problem of evaluating the
LVC reduces to the determination of the average Lagrangian velocity in each
subensemble (S), ${\bf V}^{L}(t;S)\equiv \left\langle {\bf v}\left[ {\bf x}%
(t),t\right] \right\rangle _{S}$. This is one of the advantages brought by
the subensemble analysis: the two-point LVC (\ref{CL}) can be expressed as a
function of one-point averages (${\bf V}^{L}(t;S)$) in subensembles which
correspond to given initial velocity: 
\begin{equation}
L_{ij}(t)=\int \int d\phi ^{0}\,d{\bf v}^{0}\,P_{1}(S)\;v_{i}^{0}%
\,V_{j}^{L}(t;S)  \label{3}
\end{equation}
where $P_{1}(S)\,$ is the one-point Gaussian probability density for the
values of the potential and velocity in the point $\left( {\bf 0},0\right) ,$
defined by $P_{1}(S)\,=$ $\left\langle \delta \left[ \phi ^{0}-\phi ({\bf 0}%
,0)\right] \;\delta \left[ {\bf v}^{0}-{\bf v}({\bf 0},0)\right]
\right\rangle .$ It \ represents the probability that a realization belongs
to the subensemble (S) and is obtained as: 
\[
P_{1}(S)=\frac{1}{(2\pi )^{3/2}}\frac{1}{\sqrt{E({\bf 0},0)E_{1;1}({\bf 0}%
,0)E_{2;2}({\bf 0},0)}} 
\]
\begin{equation}
\times \exp \left( -\frac{(\phi ^{0})^{2}}{2E({\bf 0},0)}-\frac{%
(v_{1}^{0})^{2}}{2E_{1;1}({\bf 0},0)}-\frac{(v_{2}^{0})^{2}}{2E_{2;2}({\bf 0}%
,0)}\right)  \label{p1s}
\end{equation}
because the potential and the velocity components are not correlated in the
same point ($\left\langle \phi ({\bf 0},0)v_{i}({\bf 0},0)\,\right\rangle =0$
when $E({\bf x})\,$has a maximum in ${\bf x=0}$, as happens in most cases).
Eq. (\ref{3}) is an exact equation.

The approximation introduced in the decorrelation trajectory method is in
the estimation of ${\bf V}^{L}(t;S)$ and essentially consists in neglecting
the fluctuations of the trajectories around the average trajectory in (S).
Thus the average Lagrangian velocity in (S) is approximated with the average
Eulerian velocity calculated along the average trajectory: 
\begin{equation}
\left\langle {\bf v}\left[ {\bf x}(t),t\right] \right\rangle _{S}\cong
\left\langle {\bf v}\left[ \left\langle {\bf x}(t)\right\rangle _{S},t\right]
\right\rangle _{S},  \label{aprox}
\end{equation}
i.e. 
\begin{equation}
{\bf V}^{L}(t;S)\cong {\bf V}^{E}\left[ {\bf X}(t,S){\bf ,}t;S\right]
\label{aprox1}
\end{equation}
where ${\bf X}(t,S)$ is the average trajectory in (S), ${\bf X}(t,S)\equiv
\left\langle {\bf x}(t)\right\rangle _{S}.$ Then, this average trajectory in
(S) (decorrelation trajectory) is determined from the equation 
\begin{equation}
\frac{d{\bf X}(t,S)}{dt}={\bf V}^{E}\left[ {\bf X}(t,S){\bf ,}t;S\right]
\label{eq}
\end{equation}
so that an explicit expression for the average Lagrangian velocity is
obtained. It was shown that Eq. (\ref{eq}) is of Hamiltonian type with the
Hamiltonian function equal to the average Lagrangian potential $\Phi ^{E}%
\left[ {\bf X}(t,S){\bf ,}t;S\right] .$ Thus, in the static case the average
trajectories obtained from the approximation (\ref{aprox}) are periodic and
evolve on closed paths. They provide a representation of the trapping and
lead to a subdiffusive transport for the static potential and to trapping
scaling for the asymptotic diffusion coefficient, $D_{tr}\sim K^{\gamma }$ \
with $\gamma <1,$ in time dependent stochastic potentials with slow
variation ($K>1$).

The approximation (\ref{aprox}) seems to be rather rough but, because it is
performed in the subensemble, there are two aspects which contribute to
improving its accuracy. One is due to the fact that the fluctuations of the
velocity in (S), which are the source of the trajectory fluctuations, are
zero in the starting point of the trajectories and become important only if
the trajectory reaches large enough distances. The second is related to the
fact that the trajectories in the subensemble are superdetermined. Besides
the necessary and sufficient initial condition ${\bf x}(0)={\bf 0,}$ they
have supplementary initial conditions determined by the definition (\ref{2})
of the subensembles. This reduces the differences between the trajectories
in (S) and thus the fluctuations. The first description of the trapping
process in qualitative agreement with the numerical simulations was obtained
using this approximations \cite{VSMB1}-\cite{PRE03}. However, there are
important qualitative aspects that are not obtained from this approximation.
The most evident concerns the average trajectory in the subensemble. In the
static case the trajectory in each realization (solution of (\ref{1})) is
periodic but the average of such trajectories cannot be periodic (as
obtained from (\ref{eq})) since they have different periods (distributed
around some average value which depends on (S)). Another aspect is discussed
in \cite{Com}, \cite{Repl} and concerns the average Lagrangian velocity in a
biased stochastic potential. It is thus necessary to improve the
decorrelation trajectory method by taking into account the fluctuations of
the trajectories in the subensembles. This development is also required by
the aim of this paper. The decorrelation trajectory method as presented in 
\cite{VSMB1} is not able to describe the statistics of the trajectories in
(S) nor the two-particle statistical quantities.

The analysis of the decorrelation trajectory method leads to the conclusion
that this method succeeds in describing the trajectory trapping due to the
fact that it maintains the invariance of the average Lagrangian potential in
(S). The motion described by Eqs. (\ref{1}), (\ref{vd}) has the velocity at
any moment perpendicular to the gradient of the potential in the
instantaneous position ${\bf x}(t)$ and the time variation of the Lagrangian
potential is produced only by the explicit time dependence of \ $\phi ({\bf x%
},t)$ 
\begin{equation}
\frac{d\phi \left[ {\bf x}(t),t\right] }{dt}=\frac{\partial \phi \left[ {\bf %
x}(t)\right] }{\partial x_{i}}\frac{dx_{i}}{dt}+\frac{\partial \phi \left[ 
{\bf x}(t),t\right] }{\partial t}=\frac{\partial \phi \left[ {\bf x}(t),t%
\right] }{\partial t}.  \label{tdep}
\end{equation}
Thus, for the static case where $\partial \phi /\partial t=0$ ($\tau
_{c},K\rightarrow \infty ),$ the potential is an invariant of the motion.
The trajectories are on the contour lines of the potential and the motion is
periodic. For slowly varying or large amplitude potentials such that $\tau
_{c}>\tau _{fl}$ \ (i.e. $K>1),$ $\partial \phi /\partial t$ is small and
there is an approximate invariance of the potential along the trajectory in
each realization. Trajectory trapping is essentially related to this
invariance of the Lagrangian potential.

The average Lagrangian potential in (S) is invariant in the frame of the
decorrelation trajectory method. Indeed, one obtains, by neglecting
trajectory fluctuations as in (\ref{aprox}), the average Lagrangian
potential in (S) as 
\begin{equation}
\Phi ^{L}(t;S)\equiv \left\langle \phi \left[ {\bf x}(t),t\right]
\right\rangle \cong \Phi ^{E}\left[ {\bf X}(t,S),t;S\right] .  \label{polagr}
\end{equation}
Since the Eulerian quantities ${\bf V}^{E}({\bf x},t;S)$ and $\Phi ^{E}({\bf %
x},t;S)$ are related by an equation similar to (\ref{vd}) 
\begin{equation}
V_{i}^{E}({\bf x},t;S)=\varepsilon _{ij}\frac{\partial \Phi ^{E}({\bf x},t;S)%
}{\partial x_{j}},  \label{rvpo}
\end{equation}
the time derivative of (\ref{polagr}) is zero for the static case and, due
to the definition (\ref{2}) of the subensembles (S), $\Phi ^{L}(t;S)=\Phi
^{E}({\bf 0};S)=\phi ^{0}$ at any time.

The aim of this paper is to consider the fluctuations of the trajectories in
the subensemble (S) and to study their effect. The approximations (\ref
{aprox}), (\ref{aprox}) that neglect trajectory fluctuations have to be
replaced. As suggested by the above discussions, this development must be
performed having in mind the idea of maintaining the condition of invariance
of $\Phi ^{L}(t;S).$ Moreover, the invariance of the Lagrangian potential
applies to each trajectory, i.e. in each realization. Thus, it determines,
besides the invariance of $\Phi ^{L}(t;S),$ other statistical constraints.
Actually, in the static case the probability distribution function for the
Lagrangian potential in a subensemble (S) is 
\begin{equation}
P^{S}(\phi ,t)\equiv \left\langle \delta \left\{ \phi -\phi \left[ {\bf x}(t)%
\right] \right\} \right\rangle _{S}=\delta (\phi -\phi ^{0})  \label{pdffi}
\end{equation}
because $\phi \left[ {\bf x}(t)\right] =\phi ^{0}$ for all trajectories in
(S) and for any time moment. We note that the knowledge of $P^{S}(\phi ,t)$
is not very helpful in solving the problem of determining the statistical
properties of the trajectories but it rather imposes strong conditions on
the possible approximations. For example, the hypothesis that trajectory
fluctuations in (S) are Gaussian around the average trajectory is excluded
by Eq. (\ref{pdffi}). It can be shown that such distribution is not
compatible with the $\delta $-distribution of the Lagrangian potential.

Trajectory fluctuations in the subensembles (S) are considered here in
agreement with the condition (\ref{pdffi}) by separating the realizations in
(S) in subensembles (S2) corresponding to fixed values of the second
derivatives of the potential in ${\bf x=0}$, $t=0$ 
\begin{equation}
(S2):\quad \phi _{ij}({\bf 0},0)\equiv \left. \frac{\partial ^{2}\phi ({\bf x%
},t)}{\partial x_{i}\partial x_{j}}\right| _{{\bf x}={\bf 0},t=0}=\phi
_{ij}^{0}  \label{s2}
\end{equation}
where $ij=11,12,22.$ The Langevin equation (\ref{1}) is studied in these
subensembles (S2). The average trajectory is approximately determined by
neglecting trajectory fluctuations in (S2). Pushed to the subensembles (S2),
this approximation is much more accurate than taken in (S) because the
trajectories in (S2) are much more superdetermined than those in (S): three
supplementary initial conditions (\ref{s2}) are added to the initial
conditions (\ref{2}). Moreover, the source of trajectory fluctuations, the
velocity fluctuations, has smaller amplitude in (S2) than in (S). Thus the
accuracy of this method is much increased compared to the accuracy of the
decorrelation trajectory method. But the main advantage of performing this
development consists in the fact that it provides for each subensemble (S) a
collection of trajectories instead of one decorrelation trajectory. It is
thus possible to determine the statistical properties of the trajectories in
(S) by performing averages over the subensembles (S2) contained in (S). We
determine using this nested subensemble procedure the dispersion of the
trajectories in (S) and their probability distribution function, besides the
average trajectory. The statistical evolution of the distance between two
trajectories is also studied by this method and the average, the dispersion
and the probability distribution function are determined as functions of
time.

We note that this nested subensemble approach can be further developed by
introducing subensembles defined by higher order derivatives of the
potential, (S3), (S4), \ .... This systematic expansion fulfils at each
order higher than 1 all the conditions required by the invariance of the
Lagrangian potential. It is however expected that the main statistical
properties of the stochastic trajectories in (S) are already obtained at the
second order and that the higher orders contribute with corrections to these
results. The nested subensembles (S), (S2) are considered in this study.

The explicit calculations consist of the following steps. First, the
statistical properties of the stochastic potential and velocity, reduced in
the subensemble (S2) defined by conditions (\ref{s2}) and (\ref{2}) are
derived. Namely, the probability that a realization belongs to a subensemble
(S2) and the subensemble average Eulerian velocity and potential are
determined. These average quantities lead to the equation for the average
trajectory in (S2) by neglecting the fluctuations of the trajectories in
(S2). Then the statistical properties of the trajectories and of the
distance between two trajectories in the subensemble (S) are expressed as
functions of the average trajectories in all subensembles (S2) contained in
(S). The calculations are for the static stochastic potential corresponding
to the strongest trapping. The transport in time-dependent potential is
discussed in Section VI B.

\subsection{Eulerian statistics in the subensemble (S2)}

The (one-point) probability that a realization of the stochastic potential
belongs to the subensemble (S2) contained in the subensemble (S) is defined
by 
\begin{equation}
P_{1}(S2)=\frac{\left\langle \delta \left[ \phi ^{0}-\phi ({\bf 0})\right]
\;\delta \left[ {\bf v}^{0}-{\bf v}({\bf 0})\right] \;\prod \delta \left[
\phi _{ij}^{0}-\phi _{ij}({\bf 0})\right] \right\rangle }{P_{1}(S)}
\label{p1s2}
\end{equation}
where the product is for $ij=11,12,22.$ It is calculated using the Fourier
representation of the $\delta $-functions and performing the average of the
resulting exponential of the stochastic Gaussian quantities (see e.g. \cite
{Feller}).\ One obtains after straightforward calculations: 
\begin{eqnarray}
P_{1}(S2) &=&\frac{1}{\left( 2\pi \right) ^{3/2}}\left[ E_{12;12}({\bf 0})c%
\right] ^{-1/2}\times  \label{p1s2r} \\
&&\exp \left[ -\frac{\left( \phi _{12}^{0}\right) ^{2}}{2E_{12;12}({\bf 0})}-%
\frac{a_{1}^{2}c_{2}}{2c}-\frac{a_{2}^{2}c_{1}}{2c}+\frac{a_{1}a_{2}c_{12}}{c%
}\right]  \nonumber
\end{eqnarray}
where $c_{i}$ are constants given by 
\begin{eqnarray}
c_{1} &\equiv &E_{11;11}({\bf 0})-E_{11;}^{2}({\bf 0})/E({\bf 0}),
\label{const} \\
c_{2} &\equiv &E_{22;22}({\bf 0})-E_{22;}^{2}({\bf 0})/E({\bf 0}),  \nonumber
\\
c_{12} &\equiv &E_{11;22}({\bf 0})-E_{11;}({\bf 0})E_{22;}({\bf 0})/E({\bf 0}%
)  \nonumber \\
c &=&c_{1}c_{2}-c_{12}^{2}  \nonumber
\end{eqnarray}
and $a_{i}$ are essentially the parameters $\phi _{ii}^{0}$ of the
subensemble (S2) 
\begin{eqnarray}
a_{1} &\equiv &\phi _{11}^{0}-\phi ^{0}E_{11;}({\bf 0})/E({\bf 0}),
\label{par} \\
a_{2} &\equiv &\phi _{22}^{0}-\phi ^{0}E_{22;}({\bf 0})/E({\bf 0}). 
\nonumber
\end{eqnarray}

The average Eulerian potential in the subensemble (S2), $\Phi ^{E}({\bf x}%
;S2)\equiv \left\langle \phi ({\bf x})\right\rangle _{S2},$ is determined by
the conditional average corresponding to (\ref{s2}): 
\begin{equation}
\Phi ^{E}({\bf x};S2)=
\end{equation}
\[
\frac{\left\langle \phi ({\bf x})\;\delta \left[ \phi ^{0}-\phi ({\bf 0})%
\right] \;\delta \left[ {\bf v}^{0}-{\bf v}({\bf 0})\right] \;\prod \delta %
\left[ \phi _{ij}^{0}-\phi _{ij}({\bf 0})\right] \right\rangle }{%
P_{1}(S)P_{1}(S2)} 
\]
This average is calculated by introducing the Fourier representation of the $%
\delta -$functions which leads to the average of $\phi ({\bf x})$ multiplied
with an exponential of a linear combination of $\phi ({\bf 0}),$ ${\bf v}(%
{\bf 0})$ and $\phi _{ij}({\bf 0}).$ This average is obtained as the
derivative at a parameter $a$ taken in $a=0$ of the average of the
exponential of the above linear combination with an additional term $a\phi (%
{\bf x}).$ One obtains after performing the inverse Fourier transforms 
\begin{equation}
\Phi ^{E}({\bf x};S2)=
\end{equation}
\[
-\frac{\left[ E({\bf x})\frac{\partial }{\partial \phi ^{0}}+E_{i;}({\bf x})%
\frac{\partial }{\partial \phi _{i}^{0}}+E_{ij;}({\bf x})\frac{\partial }{%
\partial \phi _{ij}^{0}}\right] P_{1}(S)P_{1}(S2)}{P_{1}(S)P_{1}(S2)} 
\]
which can be written explicitly as 
\begin{equation}
\Phi ^{E}({\bf x};S2)=\frac{E({\bf x})}{E({\bf 0})}\left[ \phi ^{0}+\right.
\label{fimf}
\end{equation}
\[
\left. \frac{a_{1}\left( E_{22;}({\bf 0})c_{12}-E_{11;}({\bf 0})c_{2}\right) 
}{c}+\frac{a_{2}\left( E_{11;}({\bf 0})c_{12}-E_{22;}({\bf 0})c_{1}\right) }{%
c}\right] + 
\]
\[
\frac{E_{2;}({\bf x})}{E_{2;2}({\bf 0})}v_{1}^{0}-\frac{E_{1;}({\bf x})}{%
E_{1;1}({\bf 0})}v_{2}^{0}+\frac{E_{12;}({\bf x})}{E_{12;12}({\bf 0})}\phi
_{12}^{0}+ 
\]
\[
\frac{E_{11;}({\bf x})\left( a_{1}c_{2}-a_{2}c_{12}\right) }{c}+\frac{%
E_{22;}({\bf x})\left( a_{2}c_{1}-a_{1}c_{12}\right) }{c}. 
\]
The subensemble (S2) average potential (\ref{fimf}) equals $\phi ^{0}$ in $%
{\bf x}={\bf 0,}$ $t=0$ and it decays to zero at large ${\bf x}$, as the
average Eulerian potential in the upper subensemble (S). Its expression is
more complicated and depends on the second order derivatives $\phi _{ij}^{0}$
that label (S2). The potential considered only in the realizations in the
subensemble (S2) is a non-stationary and non-homogeneous Gaussian field
having a space-time dependent average.

As in the whole ensemble and as in (S), the statistical properties of the
velocity field in the subensemble (S2) are deduced from those of the
potential in (S2). The velocity in the subensemble (S2) (\ref{s2}) is a
non-stationary and non-homogeneous Gaussian stochastic field{\bf \ }having a
space-time dependent average. This average Eulerian velocity is calculated
by the same procedure used for the potential (\ref{fimf}). The relation (\ref
{vd}) between velocity and potential in each realization holds between the
respective average quantities calculated in the nested subensembles. It was
obtained in the subensemble (S), Eq. (\ref{rvpo}), and it can be shown that
: 
\begin{equation}
V_{i}^{E}({\bf x};S2)=\varepsilon _{ij}\frac{\partial \Phi ^{E}({\bf x};S2)}{%
\partial x_{j}}.  \label{vfi}
\end{equation}
Thus the average velocity in the subensemble (S2) is \ divergence-free: $%
{\bf \nabla \cdot V}_{i}^{E}({\bf x};S2)=0.$

It is interesting to note that the potential and the velocity in the
subensembles (S) and (S2) are deterministic quantities in ${\bf x}={\bf 0}$ (%
$\phi ({\bf 0})=\phi ^{0},$ ${\bf v}({\bf 0})={\bf v}^{0}$ for all
realizations in (S), thus also in (S2)). As $\left| {\bf x}\right| $ grows,
the average values decay to zero and the fluctuations build up progressively
and eventually become the same as in the global statistical ensemble. The
amplitude of the fluctuations in a point ${\bf x}$ is smaller in the
subensemble (S2) than in the subensemble (S).

This nested subensemble procedure evidences, in the zero-average{\em \ }
stochastic velocity field, {\it a set of average velocities}{\bf \ }%
(corresponding to each subensemble). They depend on the statistical
characteristics of the velocity field (the correlation and the constraint
imposed in the problem, i.e. the zero-divergence condition). In the nested
subensembles introduced here, the following relation holds between the (S2)
average velocities and the (S) average velocity: 
\begin{equation}
{\bf V}^{E}({\bf x};S)=\int d\phi _{11}^{0}d\phi _{12}^{0}d\phi
_{22}^{0}P_{1}(S2){\bf V}^{E}({\bf x};S2).  \label{vnest}
\end{equation}
Similar equations can be written for all statistical quantities defined in
the nested subensembles.

\subsection{Average Lagrangian velocity in the subensemble (S2)}

The average Eulerian velocity (\ref{vfi}) determines an average displacement
in the subensemble (S2), ${\bf X}(t;S2).$ It is the time integral of the
average Lagrangian velocity in (S2), ${\bf V}^{L}(t;S2)$. The latter is
evaluated using an approximation similar to (\ref{aprox}) as 
\begin{equation}
{\bf V}^{L}(t;S2)\cong {\bf V}^{E}\left[ {\bf X}(t;S2);S2\right]
\label{vls2}
\end{equation}
by neglecting the fluctuations of the trajectories in the subensemble (S2).
A nonlinear equation for ${\bf X}(t;S2)$ is so obtained 
\begin{equation}
\frac{d{\bf X}(t,S2)}{dt}={\bf V}^{E}\left[ {\bf X}(t;S2);S2\right] .
\label{xms2}
\end{equation}
With this approximation the average Lagrangian \ potential in (S2) is $%
\left\langle \varphi \left[ {\bf x}(t)\right] \right\rangle _{S2}\cong \Phi
^{E}\left[ {\bf X}(t;S2);S2\right] $ and due to Eq.(\ref{vfi}) 
\[
{\bf V}^{E}\left[ {\bf X}(t;S2);S2\right] =\varepsilon _{ij}\frac{\partial }{%
\partial X_{j}}\Phi ^{E}\left[ {\bf X}(t;S2);S2\right] . 
\]
which shows that Eq. (\ref{xms2}) has a Hamiltonian structure. The solution
of this equation with the initial condition ${\bf X(}0;S2)={\bf 0}$ ensures
the invariance of its time-independent Hamiltonian function $\Phi ^{E}\left[ 
{\bf X}(t;S2);S\right] $. Since the Eulerian average potential (\ref{fimf})
has the value $\phi ^{0}$ in ${\bf x=0}$, the average Lagrangian potential
equals $\phi ^{0}$\ at any time moment.

\subsection{Lagrangian statistics in the subensemble (S)}

The nested subensemble method provides for a subensemble (S) an ensemble of
trajectories, the average trajectories ${\bf X}(t;S2),$ one for each
subensemble (S2) contained in (S). The statistical properties of these
trajectories can be determined performing averages over the subensembles
(S2).

The average velocity in (S), ${\bf V}^{L}(t;S),$ is determined by averaging $%
{\bf V}^{L}(t;S2)$ obtained from the solution of Eq.(\ref{xms2}) over all
subensembles (S2) contained in (S) 
\begin{equation}
{\bf V}^{L}(t;S)=\int d\phi _{11}^{0}d\phi _{12}^{0}d\phi _{22}^{0}P_{1}(S2)%
{\bf V}^{E}\left[ {\bf X}(t;S2);S2\right] .  \label{vls}
\end{equation}
Similar equations hold for all the statistical quantities corresponding to
the subensemble (S). The average trajectory in the subensemble (S), ${\bf X}%
(t;S)\equiv \left\langle {\bf x}(t)\right\rangle _{S},$ is determined by
averaging ${\bf X(}t;S2),$ the solution of Eq. (\ref{xms2})

\begin{equation}
{\bf X}(t;S)=\int d\phi _{11}^{0}d\phi _{12}^{0}d\phi _{22}^{0}P_{1}(S2){\bf %
X}(t;S2).  \label{avs}
\end{equation}
The dispersion of the trajectories, $d_{i}(t;S)\equiv \left\langle \left(
x_{i}(t)-{\bf X}(t;S)\right) ^{2}\right\rangle _{S}$ is 
\begin{equation}
d_{i}(t;S)=\int d\phi _{11}^{0}d\phi _{12}^{0}d\phi
_{22}^{0}P_{1}(S2)X_{i}^{2}(t;S2)-X_{i}^{2}(t;S).  \label{disps}
\end{equation}
The probability distribution function (pdf) for the trajectories in (S) is
determined by integrating the pdf in the subensemble (S2) which in this
approximation is $\delta \left[ {\bf x}-{\bf X}(t;S2)\right] $%
\begin{equation}
P^{S}({\bf x},t)=\int d\phi _{11}^{0}d\phi _{12}^{0}d\phi
_{22}^{0}P_{1}(S2)\delta \left[ {\bf x}-{\bf X}(t;S2)\right] .  \label{pdfs}
\end{equation}

The pdf for the Lagrangian potential $\phi \left[ {\bf x}(t)\right] $ in the
subensemble (S) obtained by this method equals $\delta (\phi -\phi ^{0})$
since the potential is equal to $\phi ^{0}$ on all trajectories considered
in the average. This shows that the approximation (\ref{vls2}) introduced in
this method ensures entirely the statistical properties of the Lagrangian
potential.

The statistical properties in the subensemble (S) of the distance between
two trajectories$\ {\bf \delta }(t)\equiv {\bf x}^{\prime }(t)-{\bf x}(t)$
starting from $\ {\bf x}^{\prime }(0)={\bf \delta }_{0}$ and ${\bf x}(0){\bf %
=0}$ respectively can also be determined using the average over the
subensembles (S2). The average trajectory in (S2), $\left\langle {\bf x}%
^{\prime }(t)\right\rangle _{S2},$ is the solution of Eq. (\ref{xms2}) with
the initial condition ${\bf X}^{\prime }(0;S2)={\bf \delta }_{0}.$ The
average, the dispersion and the pdf of ${\bf \delta }(t)$ in the subensemble
(S) are determined by equations similar to (\ref{avs})-(\ref{pdfs}) where $%
{\bf X}(t;S2)$ is replaced by ${\bf X}^{\prime }(t;S2)-{\bf X}(t;S2).$

\subsection{Running diffusion coefficient}

The correlation of the Lagrangian velocity in the whole set of realizations
is determined using Eq. (\ref{3}) where ${\bf V}^{L}(t;S)$ is the time
derivative of the average trajectory in (S), given by Eq. (\ref{avs}). The
running diffusion coefficient (\ref{D}) is obtained from (\ref{3}) and (\ref
{avs}) as 
\begin{equation}
D_{i}(t)=\int \int d\phi ^{0}\,d{\bf v}^{0}\,P_{1}(S)\;v_{i}^{0}\,X_{i}(t;S).
\label{ddt}
\end{equation}
In the case of an isotropic stochastic field, the integral over the
orientation of the velocity ${\bf v}^{0}$ can be performed analytically \cite
{VSMB1} and one obtains for the static case 
\begin{equation}
D(t)=\frac{1}{\sqrt{2\pi }}\frac{1}{\sqrt{E({\bf 0},0)}E_{1;1}({\bf 0},0)}%
\times  \label{dt}
\end{equation}
\[
\int_{0}^{\infty }\!d\phi ^{0}\int_{0}^{\infty }\!du\,u^{2}\exp \left( -%
\frac{(\phi ^{0})^{2}}{2E({\bf 0},0)}-\frac{u^{2}}{2E_{1;1}({\bf 0},0)}%
\right) X_{1}(t;S). 
\]
and

\begin{equation}
L(t)=D^{\prime }(t)\,,  \label{L}
\end{equation}
where $X_{1}(t;S)$ is the component of the average trajectory along ${\bf v}%
^{0},$ determined from Eq.(\ref{avs}), and $D^{\prime }(t)$ is the
derivative of the function $D(t)$. We note that the same analytical
expression for $D(t)\,$ in terms of the average trajectory $X_{1}(t;S)$ is
obtained in \cite{VSMB1} by means of the decorrelation trajectory method.
But the average trajectory was determined there as solution of a Hamiltonian
equation while here it is the average (\ref{avs}) of the average
trajectories in (S2).

\section{Explicit calculations}

The nested subensemble method actually is based on the classification of the
stochastic trajectories in groups (subensembles) according to some
resemblance between them. The most important criterion in this
classification is the value of the potential in the starting point of the
trajectories. All trajectories contained in such a group evolve on contour
lines with the same value of the potential. Consequently their paths and
periods are statistically similar in the sense that they have an average
size and period. Thus this condition determines a global resemblance of the
trajectories (extended at long time). This condition is imposed beginning
with the first level of classification (in the subensembles (S)). Other
criteria of the classification are the velocity and the derivatives of the
velocity in the origin. These are not conserved quantities ant they
influence the shape of the trajectory only at small time for time intervals
that grow with the number of imposed conditions. The value of the initial
velocity is fixed in the subensemble (S), then each subensemble (S) is
divided in smaller subensembles (S2) according to the value of the
derivatives of the velocity (second derivatives of the potential). This
classification can continue in principle and at each step the resemblance of
the trajectories contained in a group is increased and the number of groups
grows. The approximation consists in neglecting the differences between the
trajectories in a group. With this condition it is possible to determine an
average trajectory for each subensemble. Thus the nested subensemble method
determines a set of trajectories ${\bf X}(t;S2)$ and a weighting factor for
each one. Then, the statistical properties of the stochastic trajectory are
obtained by performing averages over these trajectories.

Except for some special case, the trajectories ${\bf X}(t;S2)$ have to be
numerically calculated by solving the Hamiltonian system (\ref{xms2}). This
procedure appears to be very similar with a direct numerical study of the
simulated trajectories. There are however essential differences. The average
trajectories are obtained from a rather smooth and simple time-independent
Hamiltonian. They are periodic functions and thus are calculated only for a
period. The number of trajectories is much smaller than in the numerical
study due to the weighting factor determined analytically. This reduced very
much the calculation time, such that it can be performed on PC. Moreover
such a calculation performed for a static stochastic potential with a given
EC determines the solution for the time dependent potential with arbitrary
time factor in the EC (see Section VI B).

We have developed an algorithm for calculating the statistical
characteristics of the trajectories in subensembles (S) and the running
diffusion coefficient (\ref{dt}) for given EC\ of the potential.\ The
trajectories are calculated for a period using a variable integration step
determined monitored by the precision obtained for the potential. Values of
this precision of $10^{-3}-10^{-4}$ ensure the stability of the calculated $%
D(t).$ The order of performing the integrals in Eq.(\ref{dt}) appears to be
important. The integral over $u$ is first calculated. This parameter is
factorized in the expression of the average Eulerian potential in (S2) (\ref
{fimf}) such that it appears as a factor in the right hand side of Eq.(\ref
{xms2}). This factor is introduced in the time variable and so only the
trajectories with $u=1$ need to be calculated. The values of the function $%
X_{1}(ut;S2)$ are determined by interpolation and using the periodicity. the
range of $u$ is determined from the range of the exponential factor and the
step $du$ is determined at each integration such that a large enough number
of points (30-50) exists on each period. When there are more than about 50
periods on the range of $u$ the integration is not performed because its
value is negligible. The next integrations are over $\phi _{12}^{0},$ $\phi
_{11}^{0}$ and $\phi _{12}^{0}.$ The integrand for each of these integrals
is over the range determined by the exponential factor and we have
calculated it on a mesh with constant step with 31-61 points. The last is
performed the integral over $\phi ^{0}.$ Due to trajectory trapping, the
range of this integral is continuously reduced as time increases ($\phi
_{\max }^{0}\rightarrow 0$ when $t\rightarrow \infty )$. The range of this
integral is calculated as a function of time for the interval of calculation
of $D(t).$ The calculations start with a large value of $\phi _{\max }^{0},$
as obtained from the exponential factor. At the time when the integrand
becomes approximately zero on half of this range $\phi _{\max }^{0}$ is
reduced and the integration of the trajectories is taken again from $t=0$
for the new values of $\phi ^{0}.$ The mesh for $\phi ^{0\prime }$ has
variable steps that increase toward large values of $\phi ^{0}$ because the
function has strong variations at small $\phi ^{0}.$ The tests performed
with this code have shown that the numerical calculations are rather fast
and accurate and they can be advanced up to large values of time. For
instance, using the decorrelation trajectory method the duration of the
calculation of $D(t)$ up to time of the order 10\symbol{94}2 is of the order
of 10 seconds on an usual PC. Using the nested subensemble method the
calculation time is of the order of one day because the number of calculated
trajectories increases with a factor $10^{4}$.

The nested subensemble method determines the LVC for test particles moving
in a stochastic potential with given EC. The main condition for using this
method is that the transport is stationary, which usually corresponds to
stationary and homogeneous stochastic potentials. The potential field is
Gaussian. The time-dependent diffusion coefficient (\ref{dt}) corresponds to
an isotropic potential but this is not a restriction for this method. The
above calculation are for a static potential but they can be extended to
time dependent case (see Section VI B and \cite{VSMB1}).

\section{Trajectory structures}

We present in this Section typical results obtained for the statistical
characteristics of the trajectories in a subensemble (S2). We need to
specify the EC of the stochastic potential, which we choose as 
\begin{equation}
E({\bf x})=\frac{1}{1+(x_{1}^{2}+x_{2}^{2})/2}.  \label{ecex}
\end{equation}
This is the EC of a normalized stochastic potential with amplitude $E({\bf 0}%
)=1,$ $\lambda _{c}=1$ and the time with $\tau _{fl}=1.$ The velocity ${\bf v%
}^{0}$ that defines the subensembles (S) is taken along the $x_{1}$ axis.

The average trajectory in the subensembles (S2), solution of Eq.(\ref{xms2}%
), is a periodic function of time and evolves on a closed path for most of
the subensembles (S2). There are also some open paths for subensembles with $%
\phi ^{0}=0$ and trajectories with very large periods for small values of $%
\left| \phi ^{0}\right| .$ Some examples of paths of the average
trajectories in (S2) are presented in Fig. 1. There is a clear difference
between the trajectories corresponding to small $\left| \phi ^{0}\right| $
(Fig. 1a for $\phi ^{0}=0)$ and large $\left| \phi ^{0}\right| $ (Fig. 1b
for $\phi ^{0}=1).$ In the first case there are open trajectories, large
displacements and large periods for the closed paths. In the second case the
trajectories have small size and their periods are much smaller. The size of
the path and the period of the trajectory depend on the six parameters that
define the nested subensembles (S), (S2).

The average trajectory in the upper subensemble (S) is obtained from (\ref
{avs}). Typical average trajectories in subensemble (S) are presented in
Fig. 2. The time dependence of ${\bf X}(t;S)$ is presented in Fig. 3. These
trajectories are not periodic. They evolve on spiral shaped paths, except
for the subensemble with $\phi ^{0}=0$ which yields a continuous
displacement along ${\bf v}^{0}.$ The size of the paths depends on the
parameters of (S), $\phi ^{0}$ and $u\equiv \left| {\bf v}^{0}\right| :$ it
is large for small $\left| \phi ^{0}\right| $ and large $u$ and it decreases
as{\it \ }$\left| \phi ^{0}\right| $ increases. The displacement along the
initial velocity ${\bf v}^{0}$ decays to zero in a characteristic time $\tau
_{s}$ while the displacement perpendicular to ${\bf v}^{0}$ saturates at a
finite value whose sign is the same as the sign of $\phi ^{0}.$ The
saturation time $\tau _{s}$ depends on the parameters of the subensemble
(S). It increases when $\left| \phi ^{0}\right| $ decreases (as the size of
the paths) and when $\phi ^{0}\rightarrow 0$ it becomes infinite. Thus the
average trajectories in (S) obtained here are completely different from
those obtained by means of the decorrelation trajectory method \cite{VSMB1}.
The later are periodic functions of time and their paths are closed (see
Fig. 2 for comparison). This means that the fluctuations of the trajectories
have a strong influence on the average trajectory in (S). They determine the
time-saturation of the average trajectory in (S) by the mixing of the closed
periodic trajectories.

The dispersion of the trajectories in the subensemble (S) obtained from Eq. (%
\ref{disps}) is presented in Fig. 3 as a function of time for $\phi ^{0}=0$
(Fig. 3a) and $\phi ^{0}=1$ (Fig. 3b). One can see that in the first case
the dispersion continuously increases (Fig. 3a) while in the second case it
saturates after a more complicated evolution (Fig. 3b). The saturation time
is the same as for the average trajectory. The amplitude of the trajectory
fluctuations is comparable with the average displacement.

The trajectories in the subensemble (S) are Gaussian at small time $t\ll
\tau _{fl}$ but their distribution is strongly modified as time increases.
The pdf obtained from Eq. (\ref{pdfs}) is represented in Fig. 4. Important
differences can be observed between the subensembles with $\phi ^{0}\cong 0$
(Fig. 4a) and those with large $\left| \phi ^{0}\right| $ (Fig. 4b). In the
first case the pdf is symmetric around ${\bf v}^{0},$ it develops a narrow
maximum in ${\bf x=0}$ and an annulus that expands continuously in the
direction of ${\bf v}^{0}$ as time increases (Fig. 4a). The velocity of this
part of trajectories is larger than the average velocity. The path of the
average displacement is also represented in Fig. 4a. The end point of this
curve is the average position at the moment corresponding to the
representation of the pdf ($t=100\tau _{fl}).$ It is located between the two
maxima of the pdf in a region where the later is practically zero. The pdf
for subensembles with large $\left| \phi ^{0}\right| $ is completely
different. It saturates in a time $\tau _{s}$ at a function that has a
narrow maximum in ${\bf x=0}$ and extends only on a small region (with $%
x_{2}>0$ for $\phi ^{0}>0).$

Thus the statistical characteristics of the trajectories in a subensemble
(S) with a large values of $\left| \phi ^{0}\right| $ are completely
different of those corresponding to subensembles with $\phi ^{0}\cong 0.$
The average, the dispersion and the pdf of these trajectories saturate. This
shows that there is a quasi-coherent motion in such subensembles consisting
in trajectory rotation on closed paths, with localized pdf and small
saturated dispersion. The trajectories form structures similar with fluid
vortices. These structures or eddying regions are permanent in static
stochastic potentials. The saturation time $\tau _{s}$ represents the
average time necessary for the formation of the structure. The average size
of the structure is represented by the asymptotic average displacement $%
\left| {\bf X}(t;S)\right| $ at $t\gg \tau _{s}.$ The dispersion of the
trajectories in the structure is given by the asymptotic value of $%
d_{i}(t;S).$ We have found that these characteristic parameters of the
trajectory structures depend on the parameters of the subensemble (S). The
size, the dispersion and the build up time of the structures increase when $%
\left| \phi ^{0}\right| $ decreases and go to infinity at $\phi ^{0}=0.$

The existence of the trajectory structures is confirmed by the statistical
properties of the distance between two neighboring trajectories, ${\bf %
\delta }(t)\equiv {\bf x}^{\prime }(t)-{\bf x}(t)$. Typical results obtained
for the second moment of the relative displacement ${\bf \delta }(t),$ $%
\left\langle \delta _{i}^{2}(t)\right\rangle _{S},$ are presented in Fig. 5
(continuous lines) compared with the second moments of the absolute
displacements, $\left\langle x_{i}^{2}(t)\right\rangle _{S}$ (dashed lines).
In the subensembles with large $\left| \phi ^{0}\right| $ (Fig. 5b), the
evolution of $\left\langle \delta _{i}^{2}(t)\right\rangle _{S}$ shows that
it maintains long time the initial value $\delta _{0}^{2}$ and that it
reaches values comparable with the absolute displacement $\left\langle
x_{i}^{2}(t)\right\rangle _{S}$ only after a very long time (of the order of 
$100\tau _{fl})$. Thus the relative motion is strongly hindered and a very
strong clump effect appears in the subensembles with large $\left| \phi
^{0}\right| .$ There is a very strong degree of coherence of the relative
motion for these trajectories showing that they form structures. On the
contrary, the clump effect is very weak (practically absent) for the
trajectories which are not in the structures (those in the subensembles with 
$\phi ^{0}\cong 0)$. As seen in Fig. 5a, $\left\langle \delta
_{i}^{2}(t)\right\rangle _{S}$ has values comparable to those of $%
\left\langle x_{i}^{2}(t)\right\rangle _{S}$ since the first stage of the
evolution, at time much smaller $\tau _{fl}.$ The pdf of ${\bf \delta }(t)$
in a subensemble (S) shows that the relative motion is not Gaussian. In the
case of structures (large $\left| \phi ^{0}\right| ),$ the pdf remains very
localized around zero and saturates (Fig. 6b). In the case of free
trajectories ($\phi ^{0}\cong 0),$ the pdf has a more complicated shape and
extends continuously. At large times it is similar with the pdf of the
trajectories (Fig. 6a).

Thus, the trapping of the trajectories has a strong influence on the
statistical characteristics of the relative motion. It produces an anomalous
clump effect. In the absence of trapping the clump effect appears as an
exponential time dependence of the average square distance between two
trajectories, of the type $\left\langle \delta _{i}^{2}(t)\right\rangle
=\delta _{0}^{2}\exp (t/\tau _{cl}),$ where the clump characteristic time $%
\tau _{cl}$ is of the order of the diffusion time which is the flight time $%
\tau _{fl}$ at large $K$ (see \cite{clump} and the references therein or the
review paper \cite{K02}). The distance between two neighboring trajectories
remains close to the initial value during a time $\tau _{cl}$ and then $%
\left\langle \delta _{i}^{2}(t)\right\rangle $ grows rapidly and reaches a
diffusive behavior with the diffusion coefficient equal to $2D.$ The process
of trajectory trapping determines a complete change of the clump effect. It
appears only for a part of the trajectories, those contained in subensembles
(S) with large $\left| \phi ^{0}\right| ,$ and is very strong. The life time
of the clump is much larger than $\tau _{fl}$ and than the saturation time $%
\tau _{s}.$ The time evolution of the relative square displacement is much
slower. Neighboring particles have thus a coherent motion for a long time.
For the other part of the trajectories, those that move along contour lines
of the potential with $\phi ^{0}\cong 0,$ the clump effect is absent and the
relative motion become rapidly incoherent, after a time interval smaller
than $\tau _{fl}.$

\section{Transport}

\subsection{Static stochastic potential}

The Lagrangian velocity correlation and the time dependent diffusion
coefficient for the whole ensemble of trajectories are determined from Eq. (%
\ref{dt}). The integral over the parameters of the subensembles (S) of the
average displacement in (S) has to be calculated. Since $X_{1}(t;S)$ decays
to zero in a time $\tau _{s}(S),$ the trajectory structures have only a
transient contribution to the running diffusion coefficient. At times larger
that $\tau _{s}(S)$ the contribution of the subensemble (S) vanishes. As
time increases the diffusion coefficient $D(t)$ is determined by a smaller
and smaller number of trajectories, those contained in large structures with
large saturation time. The results obtained for $D(t)$ are presented in Fig.
7 where the dimensionless function $F(t)\equiv D(t)/D_{B}$ is plotted
(continuous line). The transport is subdiffusive in such static stochastic
potential. One can observe the change that appears at $t\gtrsim \tau _{fl}.$
The running diffusion coefficient begins to decrease and eventually goes to
zero. A power law decay was obtained at $t>\tau _{fl}$ as $D(t)=V$ $\lambda
_{c}\left( t/\tau _{fl}\right) ^{-0.43}$. The LVC becomes negative at this
time and after a minimum it has a long negative algebraic tail that decays
to zero.\ The positive and the negative parts of $L(t)$ have equal time
integral such that $\int_{0}^{t}L(\tau )d\tau =D(t)\rightarrow 0.$ The mean
square displacement is $\left\langle x^{2}(t)\right\rangle \sim t^{0.57}$
and thus the process is subdiffusive.

The result obtained for $D(t)$ with the decorrelation trajectory method is
also plotted in Fig. 7 (dashed line). It is surprisingly close to the result
of the nested subensemble method although the two methods yield completely
different average trajectories in the subensembles (S) (Fig. 2). This shows
that by introducing the subensembles (S2) in the nested subensemble method a
strong qualitative improvement of the statistical results in the next upper
subensemble (S) is obtained and only a small correction at the level of the
whole set of realizations. It is thus expected that the development of the
method by introducing higher order derivatives and the corresponding nested
subensembles (S3), (S4), ...would yield only small corrections for the
physically interesting results that concern the diffusion coefficient $D(t)$
and the statistical properties of the trajectory structures. This nested
subensemble method appears to be fast convergent. This is a consequence of
the fact that the mixing of periodic trajectories, which characterizes this
nonlinear stochastic process, is directly described at each order of our
approach. The results obtained in first order (the decorrelation trajectory
method) for $D(t)$ are thus validated by the present second order
calculations.

\subsection{Time-dependent stochastic potential}

For a time-dependent potential $\phi ({\bf x},t)$ (finite $\tau _{c}$ and $%
K) $, it is also possible to apply the nested subensemble method following
the same procedure as above. A very simple analytical solution is obtained
when the stochastic potential has independent time and space variations such
that its EC is $E({\bf x})h(t).$ In this case, the average Eulerian
potential in the subensemble (S) is given by Eq.(\ref{fimf}) multiplied with
the factor $h(t).$ This factor is transmitted to the average Eulerian
velocity in (S) (\ref{vfi}) and it appears in the equation (\ref{xms2}) for
the average trajectory in (S2). A change of variable from $t$ to 
\begin{equation}
\theta (t)=\int_{0}^{t}h(\tau )d\tau  \label{teta}
\end{equation}
can be performed in Eq.(\ref{xms2}) and thus the average trajectory in (S2)
for a time dependent potential can be written in terms of the average
trajectory for the static case as ${\bf X}(\theta (t);S2).$ The argument $%
\theta (t)$ determined by the time-dependence of the potential is $\theta
(t)\cong t$ at small $t$ and saturates at a constant which is the
decorrelation time $\theta (t)\rightarrow \tau _{c}.$ The same expression (%
\ref{dt}) is eventually obtained for the time dependent diffusion
coefficient but with $X_{1}(t;S)$ replaced by $X_{1}(\theta (t);S)$ and thus
the diffusion coefficient is 
\begin{equation}
D^{td}(t)=D\left[ \theta (t)\right] .  \label{dtd}
\end{equation}
The limit for $t\rightarrow \infty $ is finite which shows that the
transport is diffusive in a time dependent stochastic potential and the
diffusion coefficient is 
\begin{equation}
D^{td}=D(\tau _{c})=D_{B}F(\tau _{c}).  \label{dastd}
\end{equation}
This equation shows that the asymptotic diffusion coefficient is determined
by the time dependent diffusion coefficient $D(t)$ corresponding to the
static potential ($F(t)$ is the function plotted in Fig.7 and represents the
normalized diffusion coefficient in the static potential and $D_{B}=(\lambda
_{c}^{2}/\tau _{c})K=V\lambda _{c}$ is the Bohm diffusion coefficient
obtained when trajectory trapping is neglected). In the limit of small $K,$
the quasilinear result is recovered from Eq.(\ref{dastd}) and, at large $K,$ 
$D^{td}$ is reduced compared to the Bohm diffusion coefficient by a factor $%
F(K)<1$ which accounts for trajectory trapping. For the above EC of the
potential, Eq.(\ref{dastd}) gives the large $K$ scaling law $D^{td}\approx ($
$\lambda _{c}^{2}/\tau _{c})K^{\gamma }$ with $\gamma =0.57$. The exponent $%
\gamma $ depends on the EC of the potential, namely on its space dependence
at large distances. It is not a fixed value as in the estimation based on
percolation theory \cite{isichenko}. A detailed study of the effect of the
EC of the potential on the scaling of the diffusion coefficient is in
progress.

Thus, the study of the static case permits to determine the asymptotic
diffusion coefficient in a time dependent stochastic potential. This
property appears in the results of the nested subensemble method (and in the
decorrelation trajectory method) but it is possible to demonstrate in
general that the time dependence of the diffusion coefficient in the
subdiffusive static case determines the Kubo number dependence of the
asymptotic diffusion coefficient in a time dependent potential. The
subdiffusive transport corresponds to Lagrangian correlations $L(t)$ which
have the property: 
\begin{equation}
D(t)=\int_{0}^{t}L(t)dt\rightarrow 0.  \label{subdif}
\end{equation}
This shows that such a correlation has negative parts that compensate the
small time correlation which is always positive. We suppose that $D(t)$
decays to zero as $D(t)\approx \left( t/\tau _{fl}\right) ^{-\alpha }$ and
consequently the LVC behaves as $L(t)\approx (t/\tau _{fl})^{-\alpha -1}.$
In the time-dependent case, the variation of the stochastic field produces
the decorrelation of the Lagrangian velocity and consequently the Lagrangian
correlation decays to zero at $t\gtrapprox \tau _{c}.$ The asymptotic
diffusion coefficient can be evaluated as 
\[
D=\int_{0}^{\infty }L(t)dt\simeq \int_{0}^{\tau _{c}}L(t)dt 
\]
and using Eq. (\ref{subdif}) one can write 
\[
D\simeq -\int_{\tau _{c}}^{\infty }L(t)dt\approx (\tau _{c}/\tau
_{fl})^{-\alpha }=K^{-\alpha }. 
\]
Thus the exponent $\alpha $ of the time decay of the subdiffusive transport
coefficient in the static case determines the exponent of the $K$ dependence
of the asymptotic diffusion coefficient in the time-dependent case. This
means that Eq.(\ref{dastd}) holds even if the evolution of $D^{td}(t)$ is
not given by Eq.(\ref{dtd}) as happens for example when the EC of the
potential is not factorized.

The time variation of the potential determines a decorrelation effect. After
a time of the order $\tau _{c}$ the configuration of the stochastic
potential changes. A competition appears between the intrinsic tendency of
the trajectories to form structures and the destruction of these structures
produced by the time variation of the potential field. Structures with $\tau
_{s}(S)\gtrsim \tau _{c}$ cannot exist and the corresponding trajectories
produce a diffusive transport. Small structures that build up rapidly (with $%
\tau _{s}(S)\ll \tau _{c})$ still exist if the correlation time of the field
is longer than the flight time ($\tau _{c}>\tau _{fl},$ $K>1).$ These
vortical structures do not contribute to the large time values of the
diffusion coefficient and the transport is reduced.

\section{Conclusions}

We have studied the special problem of test particle transport in
2-dimensional divergence-free stochastic velocity fields, which is
characterized by the intrinsic trapping of the trajectories on the contour
line of the stochastic potential. We have developed a semi-analytical
statistical approach, the nested subensemble method. The time dependent
diffusion coefficient is determined by means of a set of deterministic
trajectories, the average trajectories in subensembles with given values of
the potential and of the velocity in their starting point. These
trajectories are obtained by dividing each subensemble in a class of
subensembles defined by the values of the second derivatives of the
potential. Thus, the nested subensemble approach reduces the problem of
determining the statistical behavior of the stochastic trajectories to the
calculation of weighted averages of some smooth, deterministic trajectories
determined from the EC of the stochastic potential.

The statistical characteristics of subensembles of trajectories are obtained
with this method. We have shown that the statistical behavior of the trapped
trajectories is completely different from that of the free trajectories. The
trapped trajectories have a quasi-coherent behavior. Their average
displacement, dispersion and probability distribution function saturate. A
very strong anomalous clump effect characterizes neighboring trapped
trajectories. Their clump life time is very large compared to the time of
flight. This shows that these trajectories form structures similar with
fluid vortices. The statistical parameters of these structures (size,
build-up time, dispersion) are determined. The trajectories contained in
such structures do not contribute to the large time diffusion coefficient.
The later is determined by the free trajectories which have a continuously
growing average displacement and dispersion. The probability distribution
function for both types of trajectories are non-Gaussian.

The time dependent diffusion coefficient is determined as a functional of
the Eulerian correlation of the stochastic potential.

The general conclusion of this work is that the existence of an invariant in
the evolution equation (the potential) determines long-time correlations
(memory effects) and coherence (trajectory structures) in the stochastic
motion.

\noindent {\bf Acknowledgments}

This work is performed during our stay at National Institute for Fusion
Science, Japan as visiting professors. We warmly acknowledge the hospitality
of Professor M. Fujiwara and Professor O. Motojima. We want to thank
Professor K. Itoh for very fruitful discussions and Professor R. Balescu and
Dr. J. H. Misguich for their stimulating interest in this problem.


\onecolumn



\vspace*{0.30in}

\begin{center}
\resizebox{4.in}{!}{\includegraphics{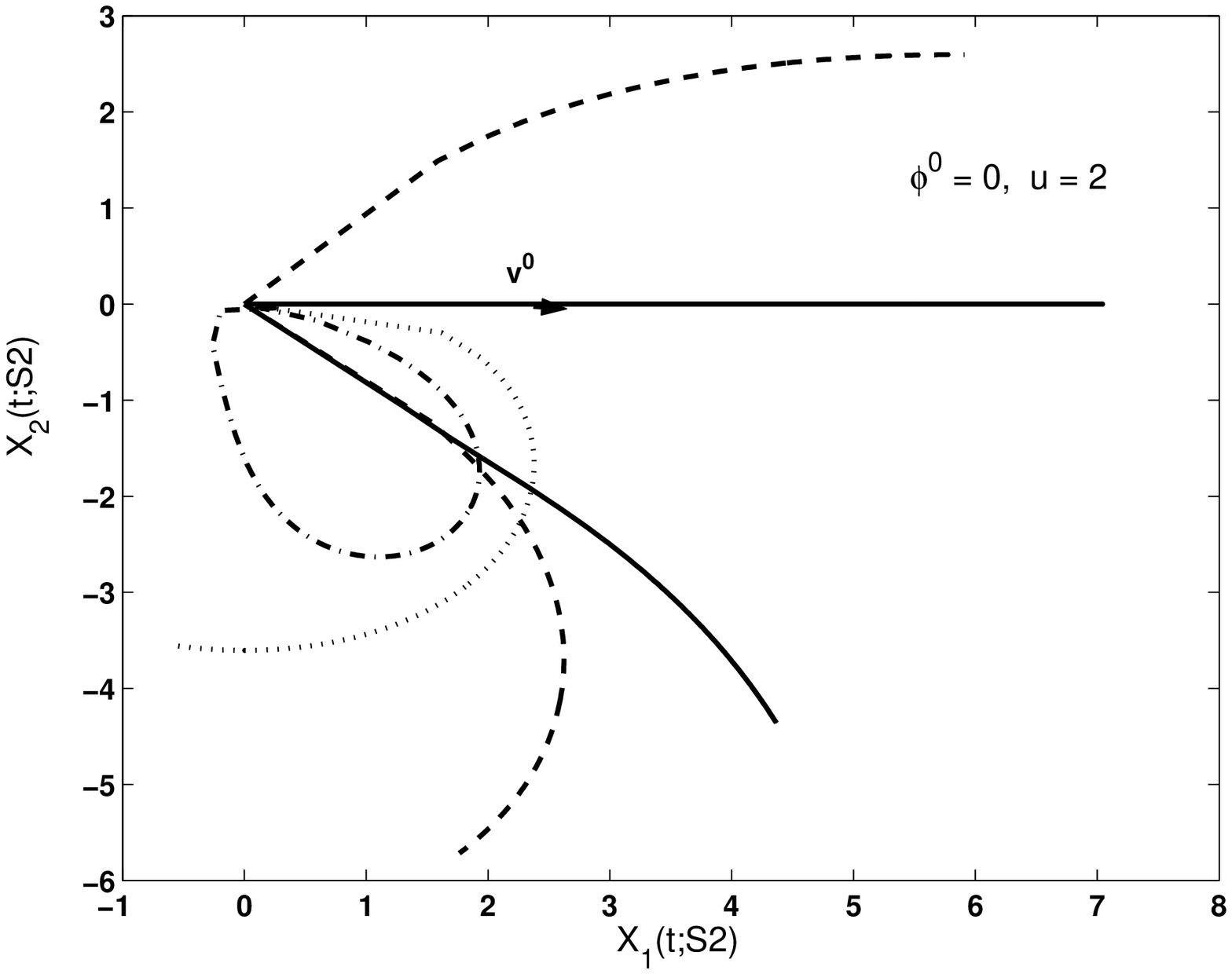}} 

\vspace*{0.10in}

a 
\end{center}



\begin{center}
\resizebox{4.in}{!}{\includegraphics{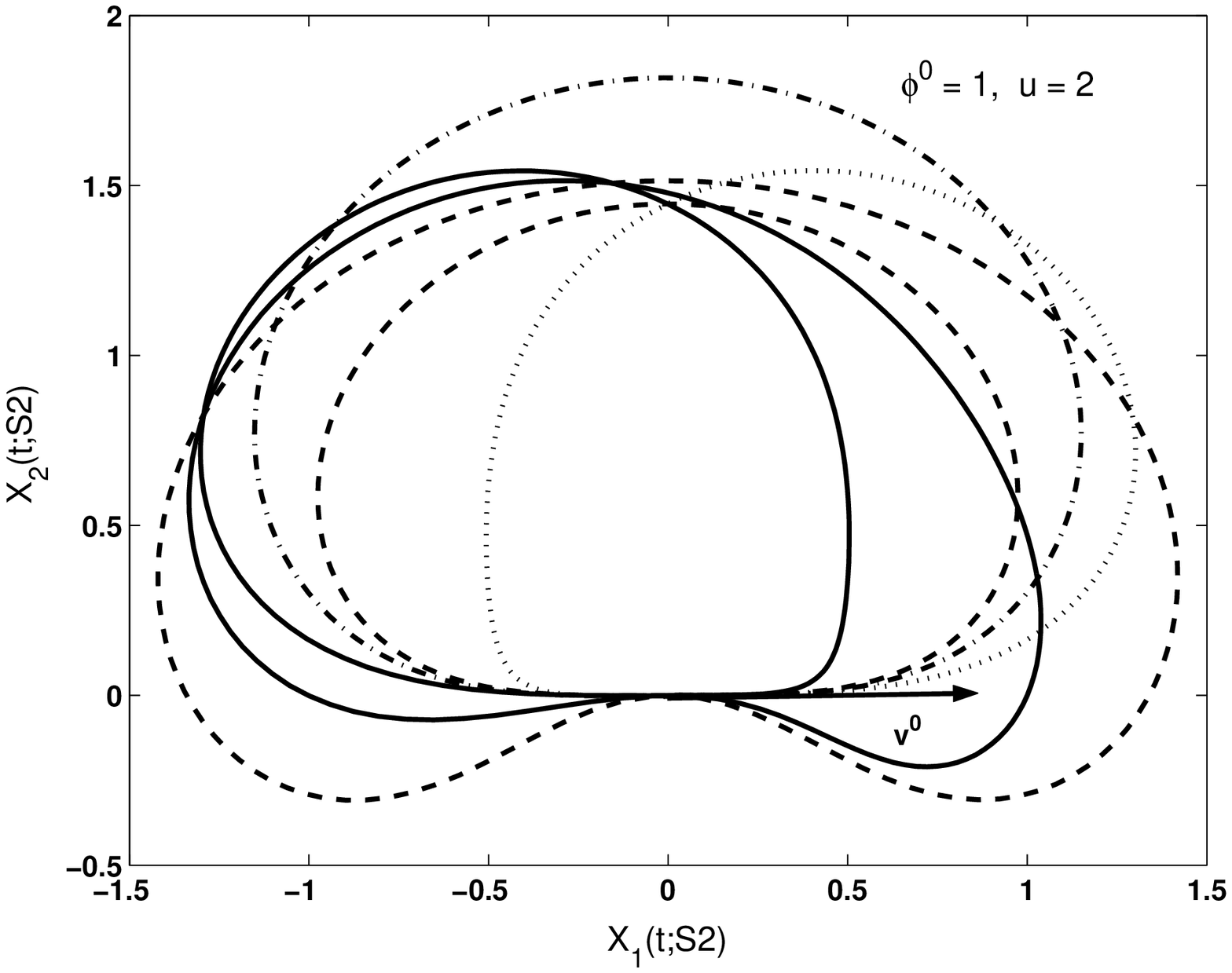}} 

\vspace*{0.10in}

b
\end{center}

\vspace*{0.30in}

\begin{center}
Figure 1 \\
Exemples of paths of average trajectories in subensembles (S2) obtained from Eq. (32) for several values of $\phi _{ij}^{0}$ and for the values of $\phi ^{0}$ and $u$ mentioned on the figure.
\end{center}

\vspace*{2.0in}

\begin{center}
\resizebox{4.in}{!}{\includegraphics{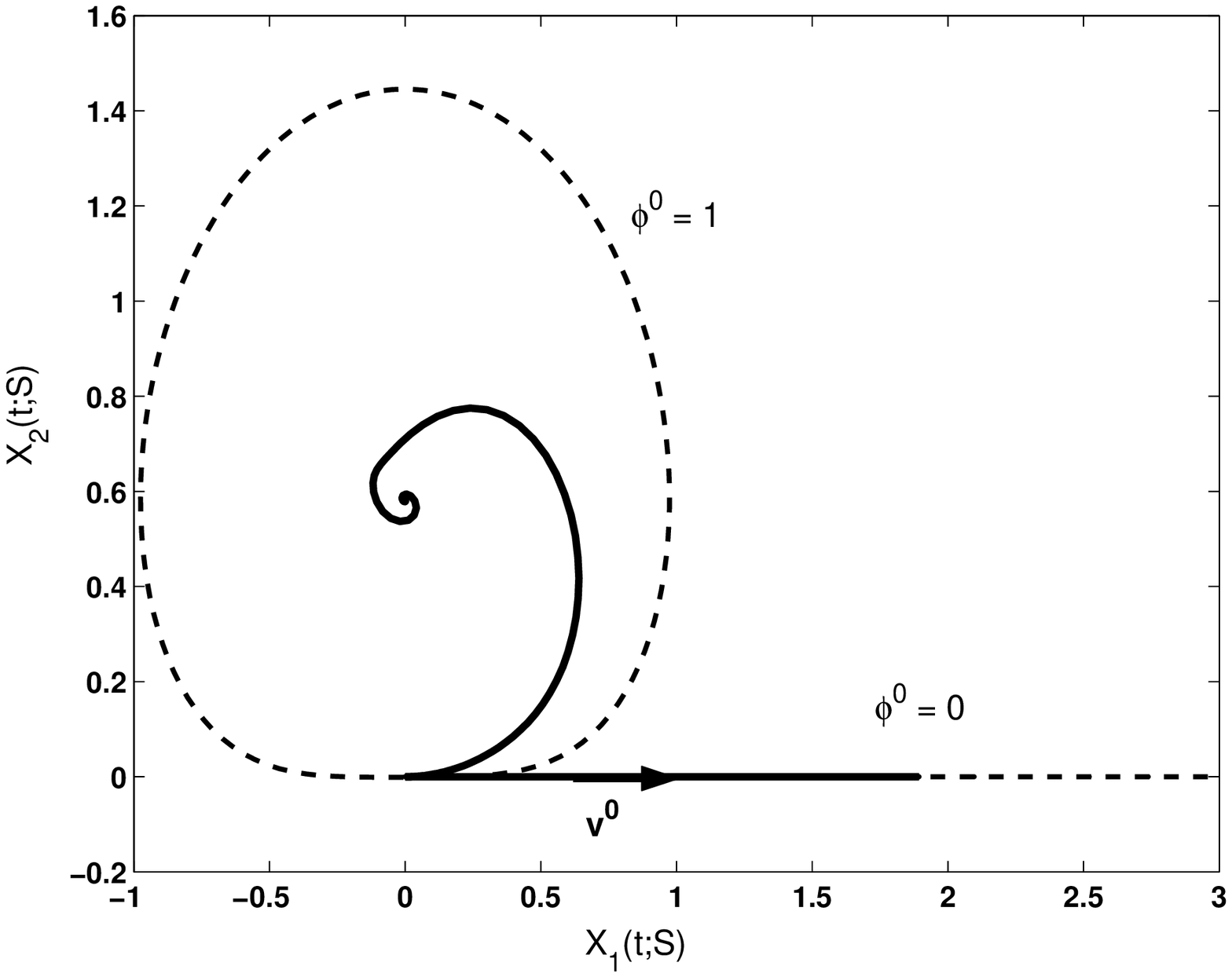}} 

\vspace*{0.30in}

Figure 2\\
Exemples of paths of average trajectories in subensembles (S) obtained from Eq. (34) (continuous lines). The results obtained with the decorrelation trajectory method are also ploted for comparision (dashed lines).
\end{center}

\newpage


\vspace*{0.3in}

\begin{center}
\resizebox{4.in}{!}{\includegraphics{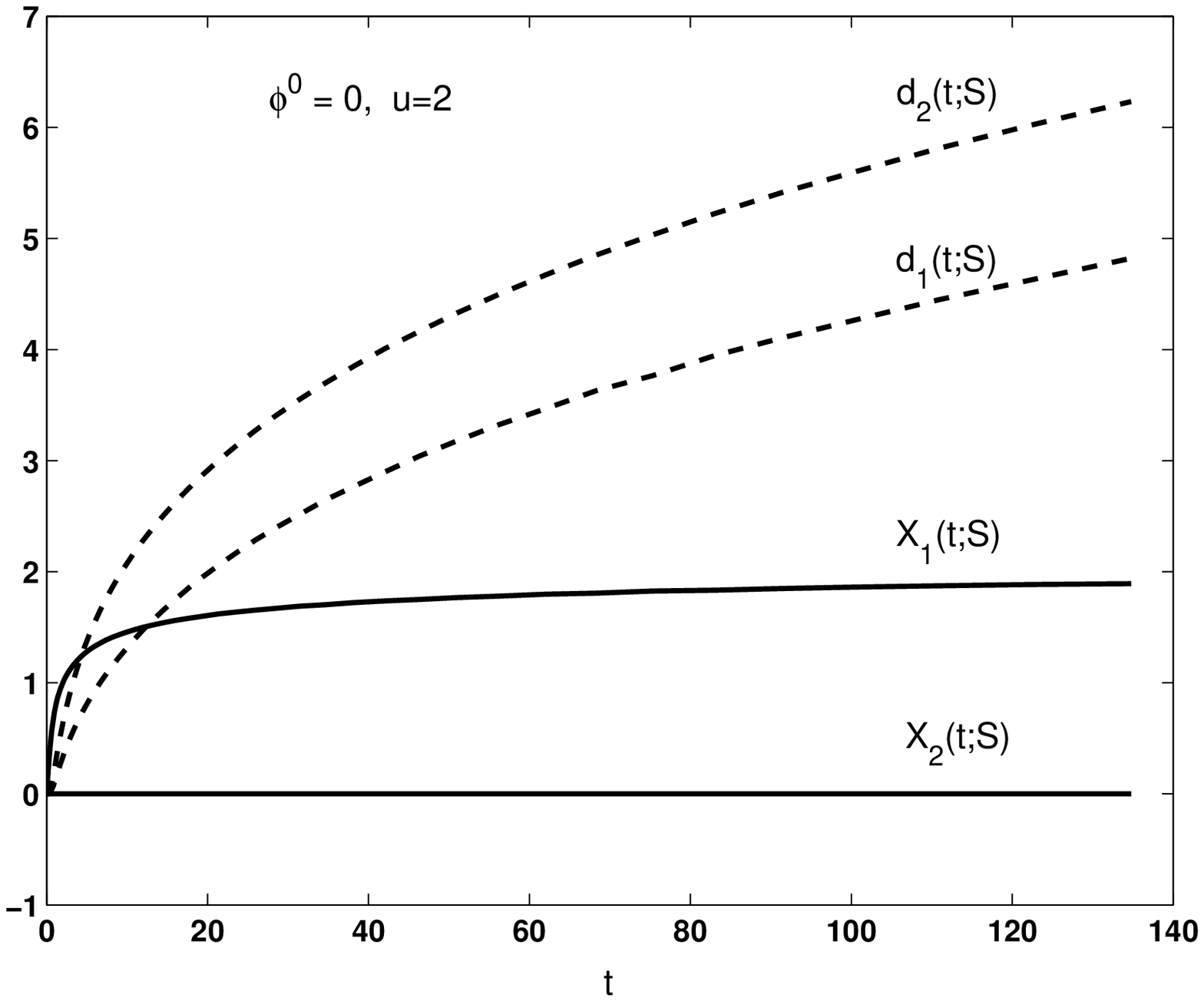}} 

\vspace*{0.10in}

a 
\end{center}




\begin{center}
\resizebox{4.in}{!}{\includegraphics{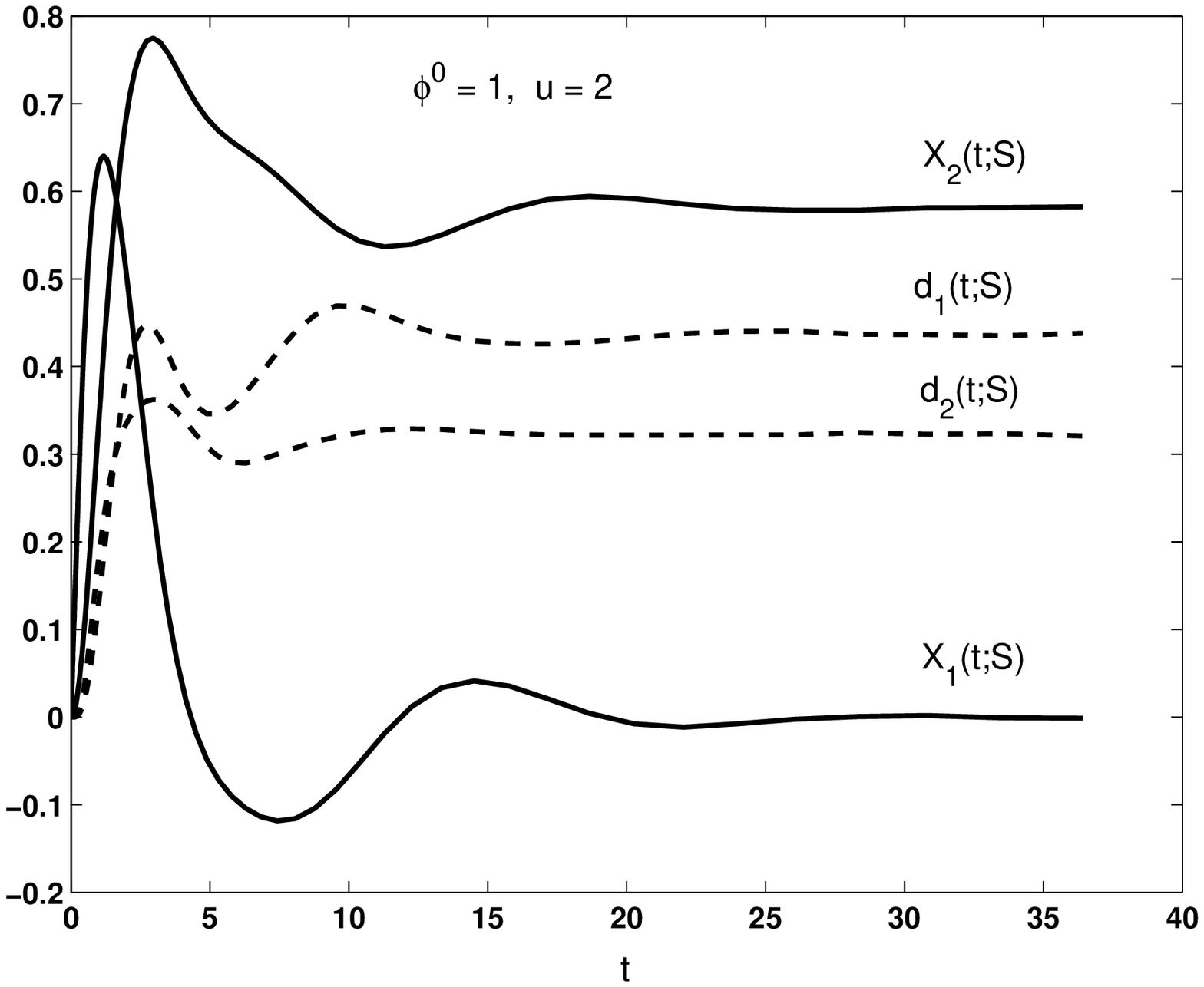}} 

\vspace*{0.10in}

b
\end{center}

\vspace{0.30in}

\begin{center}
Figure 3 \\ 
Time evolution of the average trajectory and of the dispersion [Eq. (35)]  in subensembles (S) with $\phi ^{0}=0$ (a) and for $\phi ^{0}=1$ (b).
\end{center}

\newpage


\vspace*{0.3in}

\begin{center}
\resizebox{4.in}{!}{\includegraphics{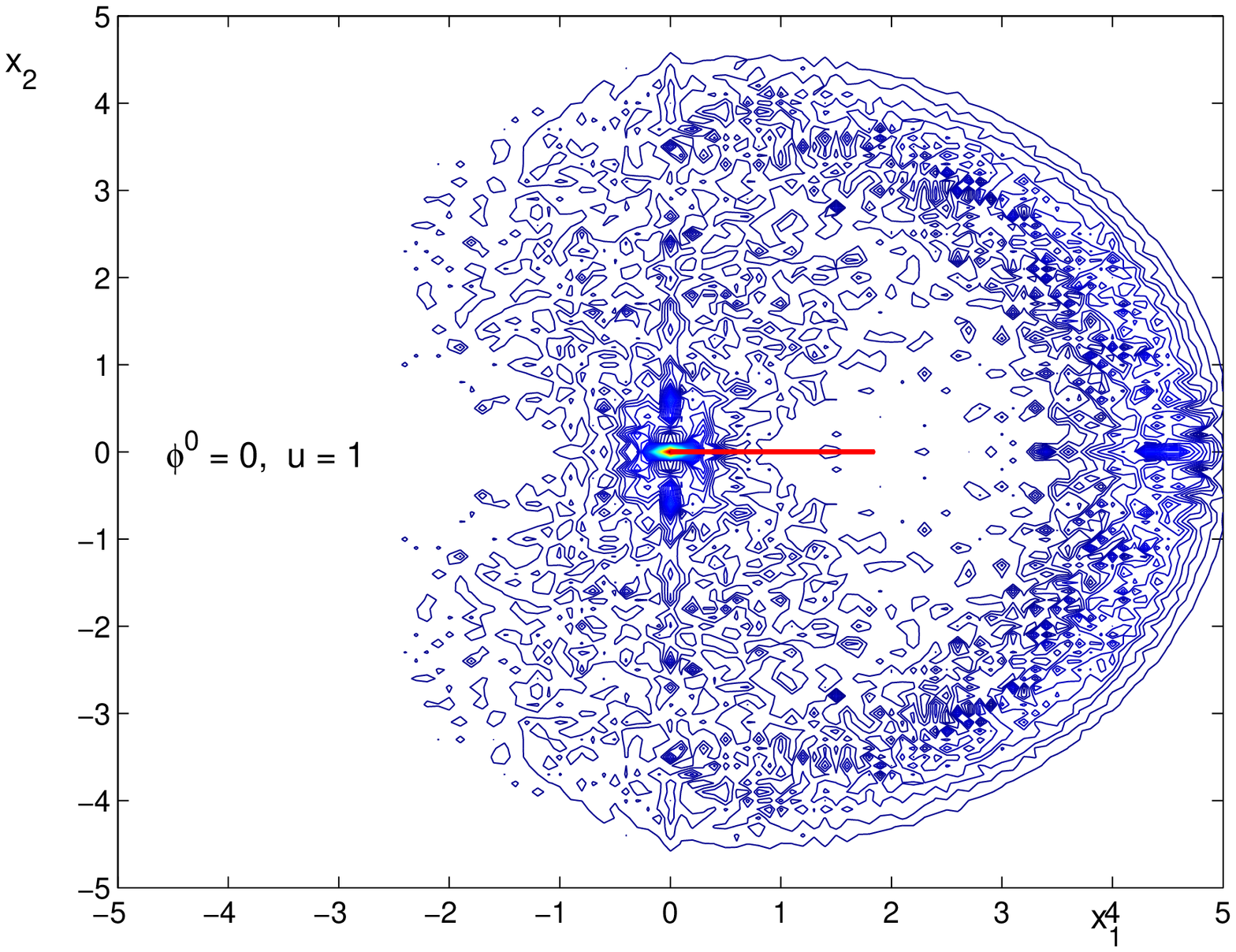}} 

\vspace*{0.10in}

a 
\end{center}




\begin{center}
\resizebox{4.in}{!}{\includegraphics{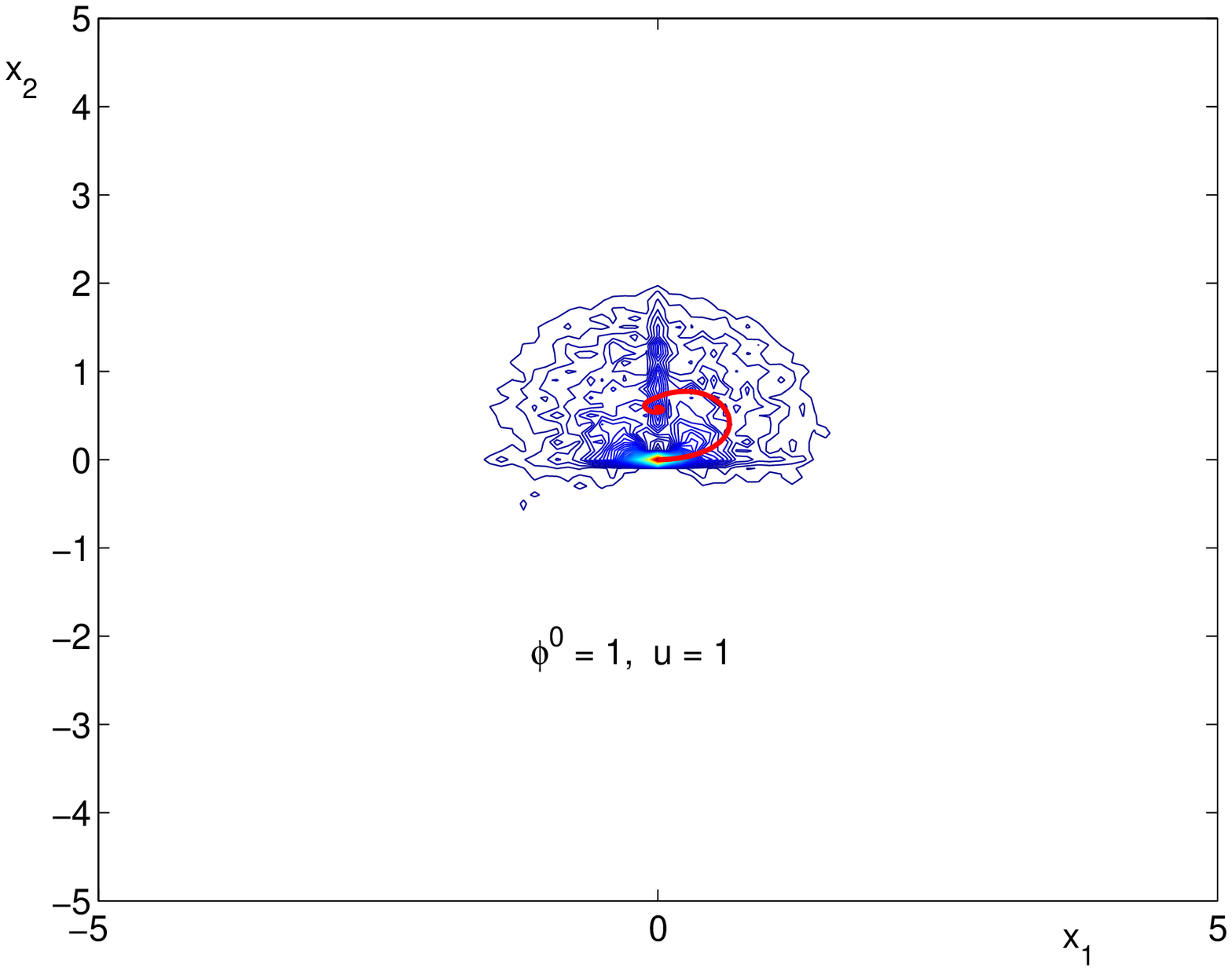}} 

\vspace*{0.10in}

b
\end{center}

\vspace{0.30in}

\begin{center}
Figure 4 \\ 
Contour plot of the pdf [Eq. (36)] of the trajectories in a subensemble (S) for (a) $\phi ^{0}=0,$ $u=1,$ $t=..\tau _{fl}$ \ and (b) $\phi ^{0}=1,$ $u=1,$ $t=..\tau _{fl}$ (at saturation). The path of the average trajectory is also represented (red line) on the interval $\left[ 0,t\right] .$
\end{center}

\newpage


\vspace*{0.3in}

\begin{center}
\resizebox{4.in}{!}{\includegraphics{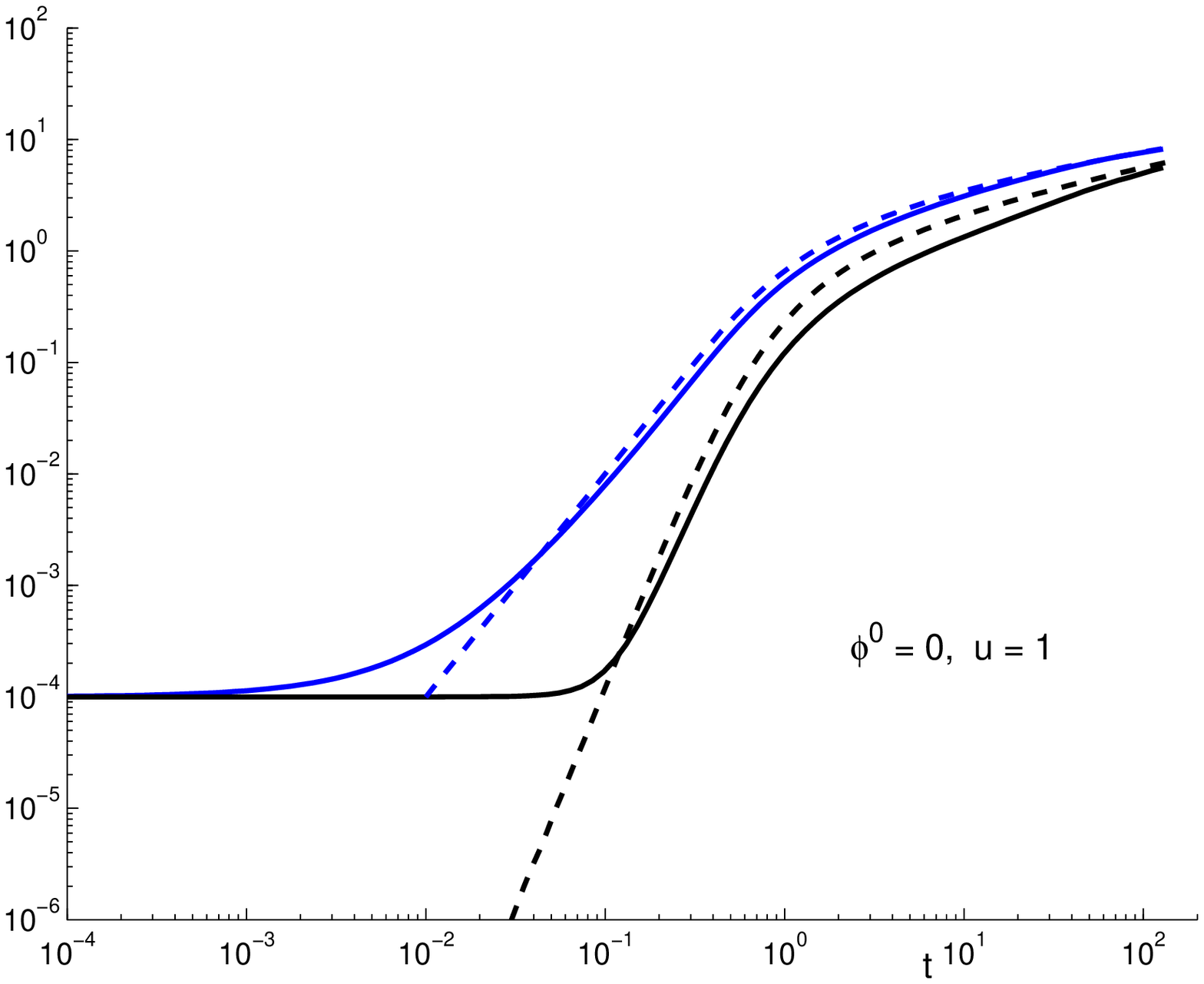}} 

\vspace*{0.10in}

a 
\end{center}




\begin{center}
\resizebox{4.in}{!}{\includegraphics{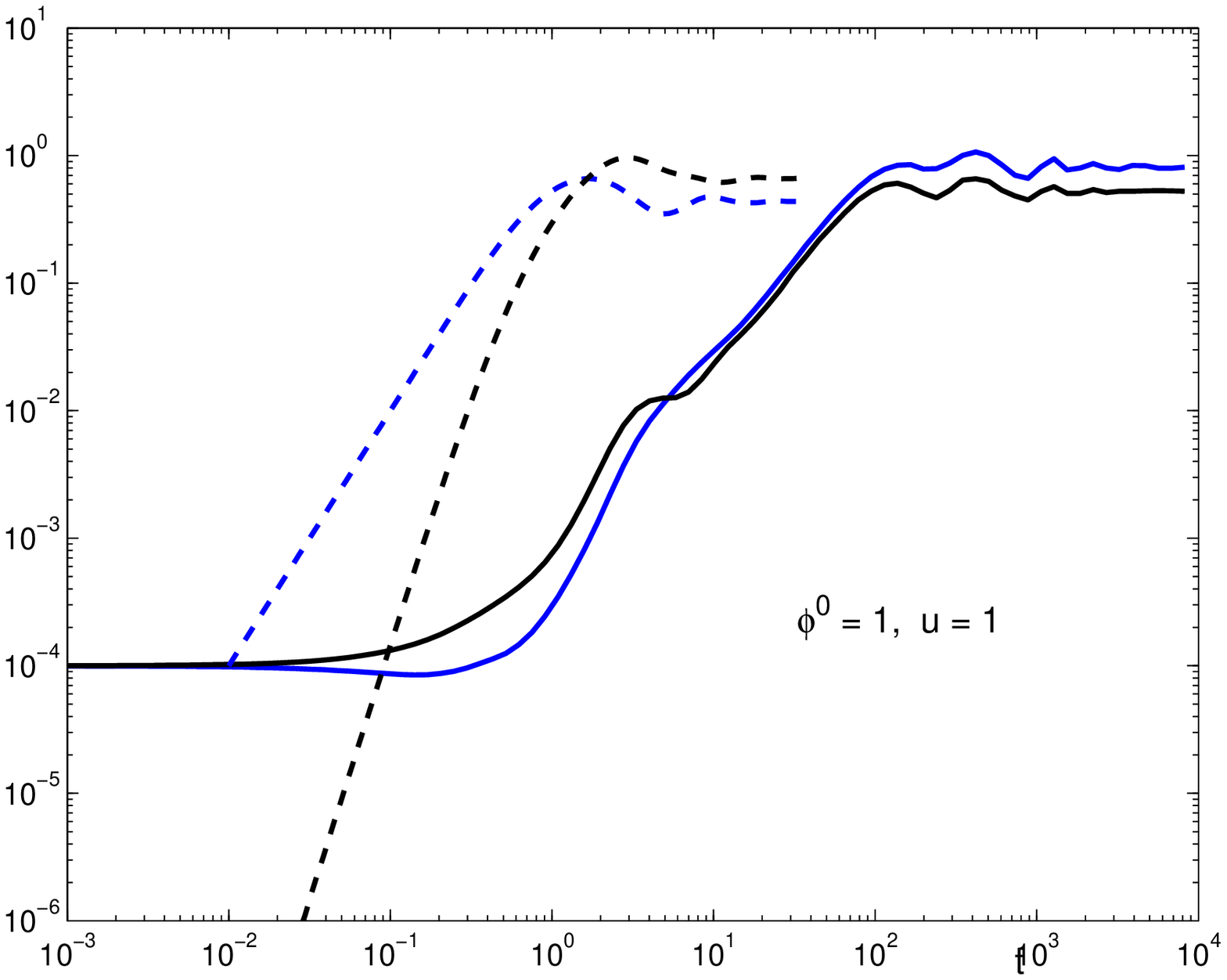}} 

\vspace*{0.10in}

b
\end{center}

\vspace{0.30in}

\begin{center}
Figure 5 \\ 
The second moments of the relative displacement $\left\langle \delta _{1}^{2}(t)\right\rangle _{S}$ (continuous blue line) and $\left\langle \delta _{2}^{2}(t)\right\rangle _{S}$ (continuous black line) compared to $\left\langle x_{1}^{2}(t)\right\rangle _{S}$ (dashed blue line) and $\left\langle x_{2}^{2}(t)\right\rangle _{S}$ (dashed black line) for two subensembles (S): one with $\phi ^{0}=0,$ $u=1$ (Fig. 5a) and the other with $\phi ^{0}=1,$ $u=1$ (Fig. 5b).
\end{center}

\newpage


\vspace*{0.3in}

\begin{center}
\resizebox{4.in}{!}{\includegraphics{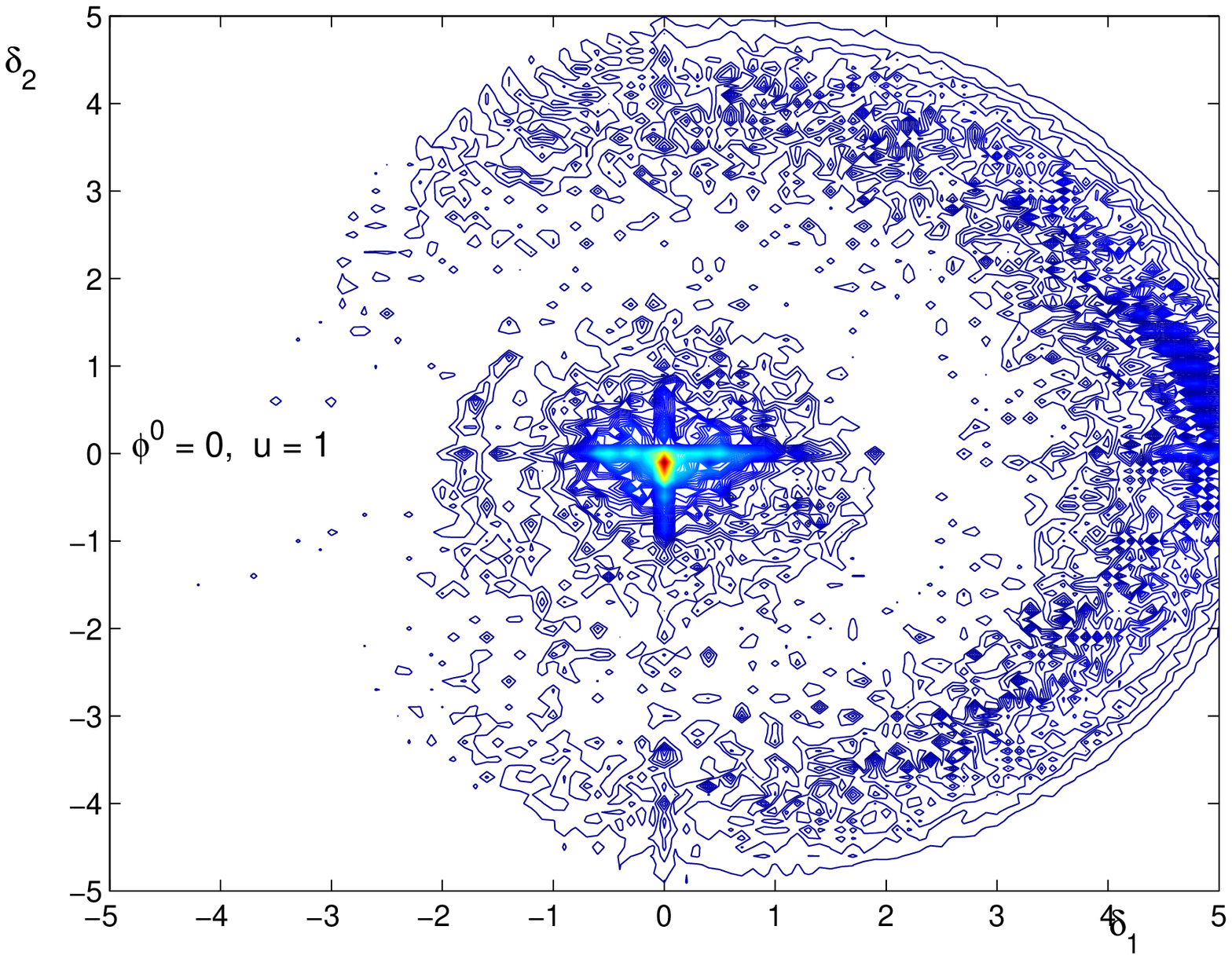}} 

\vspace*{0.10in}

a 
\end{center}




\begin{center}
\resizebox{4.in}{!}{\includegraphics{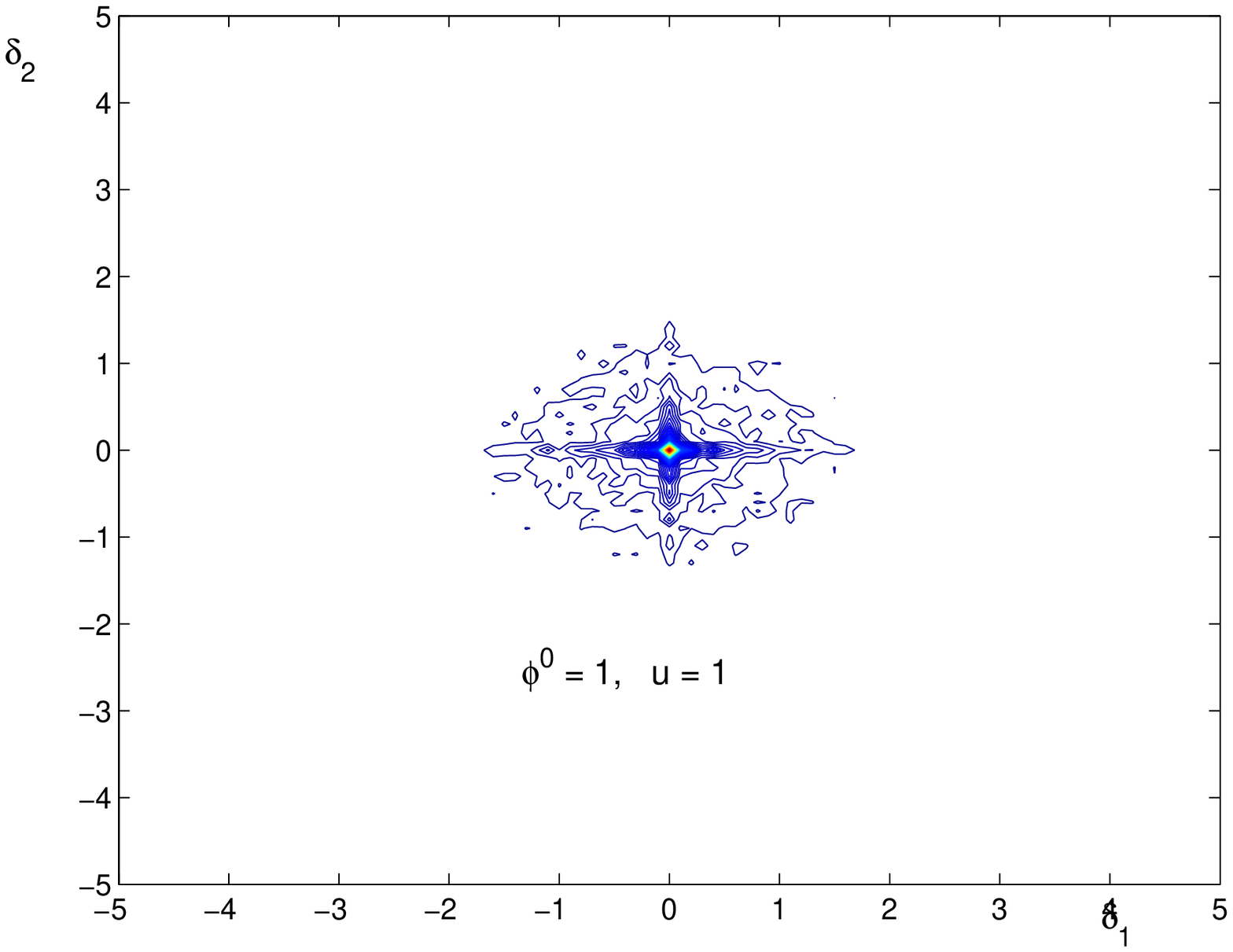}} 

\vspace*{0.10in}

b
\end{center}

\vspace{0.30in}

\begin{center}
Figure 6 \\ 
Contour plot of the pdf of the relative motion in a subensemble (S) for (a) $\phi ^{0}=0,$ $u=1,$ $t=..\tau _{fl}$ \ and (b) $\phi ^{0}=1,$ $u=1,$ $t=..\tau _{fl}$ (at saturation).
\end{center}

\newpage

\vspace*{2.0in}

\begin{center}
\resizebox{4.in}{!}{\includegraphics{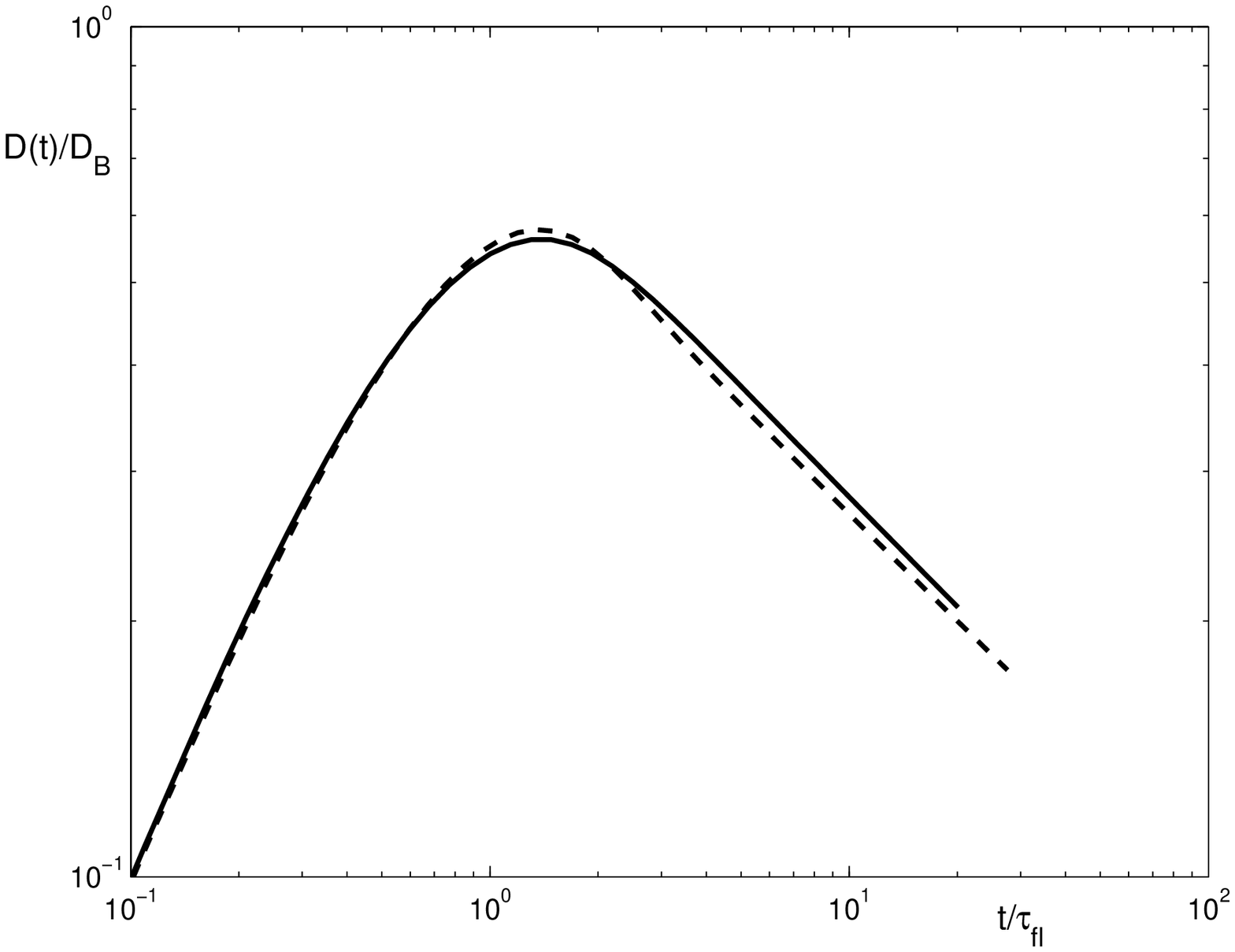}} 

\vspace*{0.30in}

Figure 7\\
The time dependent diffusion coefficienr $D(t)$ obtained from Eq. (38) with the nested subensemble method (continuous line) compared with the result of the decorrelation trajectory method (dashed line).
\end{center}

\end{document}